\documentclass{article}

\usepackage{PRIMEarxiv}

\usepackage[utf8]{inputenc} 
\usepackage[T1]{fontenc}    
\usepackage{hyperref}       
\usepackage{url}            
\usepackage{booktabs}       
\usepackage{amsfonts}       
\usepackage{nicefrac}       
\usepackage{microtype}      
\usepackage{lipsum}
\usepackage{fancyhdr}       
\usepackage{graphicx}       
\graphicspath{{media/}}     
\usepackage{amsthm}
\usepackage{colortbl} 
\usepackage{lscape}
\usepackage{tablefootnote}
\usepackage{xcolor}
\usepackage{amsmath}
\usepackage{makecell}
\newtheorem{definition}{Definition}
\newtheorem{proposition}{Proposition}
\title{Streamlining HTTP Flooding Attack Detection through Incremental Feature Selection
}

\author{
    \textbf{Upasana} Sarmah \and \textbf{Parthajit Borah} \and \textbf{D. K. Bhattacharyya} \\
    \normalsize Department of Computer Science \& Engineering, Tezpur University, Tezpur, Assam \\
    \normalsize \texttt{\{upatink, parthajit, dkb\}@tezu.ernet.in} 
}

\begin{document}
\maketitle

\begin{abstract}
Applications over the Web primarily rely on the HTTP protocol to transmit web pages to and from systems. There are a variety of application layer protocols, but among all, HTTP is the most targeted because of its versatility and ease of integration with online services. The attackers leverage the fact that by default no detection system blocks any HTTP traffic. Thus, by exploiting such characteristics of the protocol, attacks are launched against web applications. HTTP flooding attacks are one such attack in the application layer of the OSI model. In this paper, a method for the detection of such an attack is proposed. The heart of the detection method is an
incremental feature subset selection method based on mutual information and correlation. INFS-MICC helps in identifying a subset of highly relevant and independent feature subset so as to detect HTTP Flooding attacks with best possible classification performance in near-real time. 

\end{abstract}


\section{Introduction}
HTTP flooding attacks on the application layer of the OSI model are a kind of DDoS attack where the services offered by a Web server are brought down by an attacker. These kinds of attacks consume less bandwidth as the attacker floods the server with legitimate HTTP requests. After a certain point of time upon flooding, the server is overwhelmed and cannot respond to legitimate requests of its user base. The attacker customizes the requests according to the target web application and the ultimate aim is to exhaust the server's limited resources. HTTP protocol is the most targeted protocol due to its versatility and ease of use with a variety of Web applications \cite{singh2008vulnerability}. It is important to note here, that an attacker may employ a botnet army to launch a coordinated attack to bring down the services. \\
Compared to others, there is far less difference between a legitimate HTTP request and an attack request, which is why these attacks are hard to detect \cite{sreeram2019http}. This is because the underlying Transmission Control Protocol (TCP) and User Datagram Protocol (UDP) connections for both an attack request and a normal HTTP request are the same. The only difference between the two is the intent with which a request is made. Additionally, as seen from real instances of application layer DDoS attacks\footnote{\url{https://www.enisa.europa.eu/publications/enisa-threat-landscape-for-dos-attacks}} it is difficult to perceive the starting and ending point of an attack. Depending on the scale of the attack, the disruptions caused by the attack often live longer than the attack itself, and this is the reason why it is hard to estimate the actual length of an attack.\\
The selection of appropriate characteristics or features plays an important role in identifying actions that can lead to an attack. All features may not be equally informative; some may be redundant or irrelevant and thus play no role in the detection process \cite{langley1994selection}. In fact, a subset of highly informative features should be selected for good performance. On the other hand, in the case of a dynamic real-world application all the data may not be available at one time. Such cases necessitate the use of incremental learning, so as to avoid learning from scratch each time data is available \cite{luo2020appraisal}. Thus, a defense mechanism that incrementally selects relevant and irredundant features will definitely aid in solving the problem of HTTP flooding attacks in the application layer.\\

\subsection{Attack Statistics and Defense Systems in Applications}\label{real_world_httpf}
\begin{figure}[ht!]
\centering
	\includegraphics[width=0.75\linewidth]{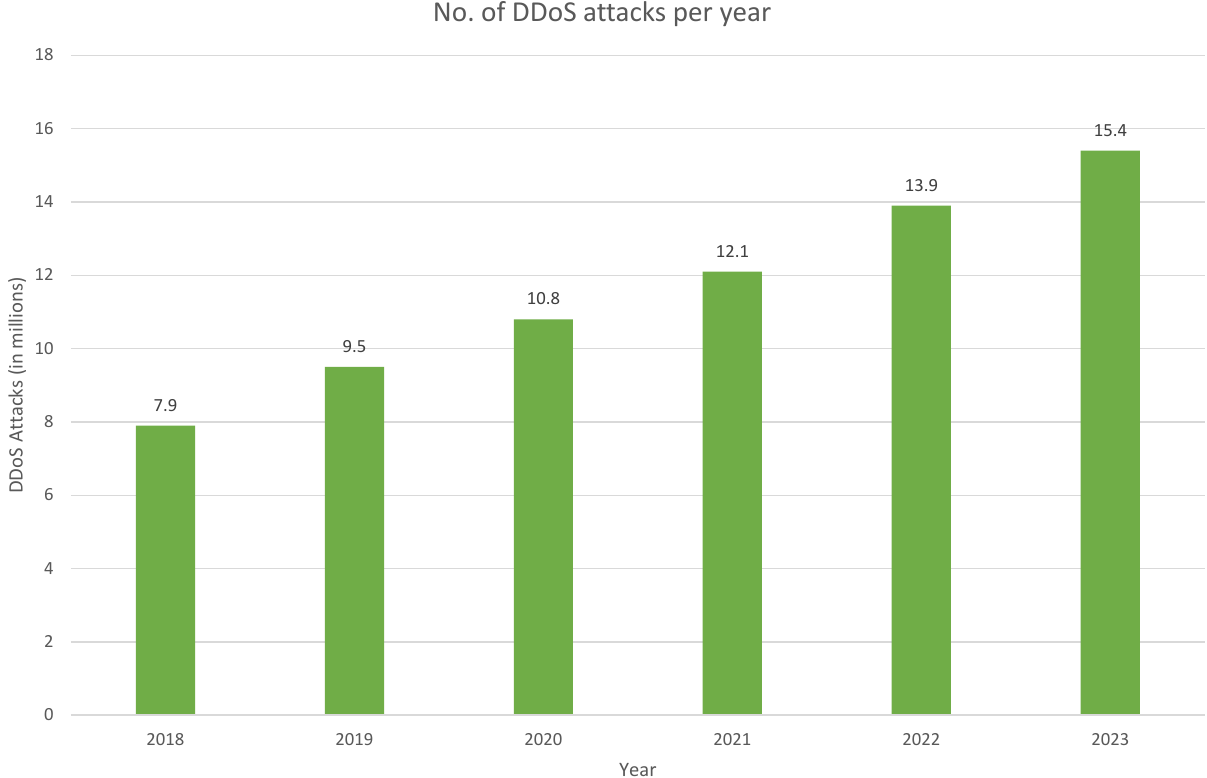}
\caption{DDoS Attacks Over the Years}
	\label{DDoS Attacks Over the Years}
\end{figure}

The Cisco Annual Report\footnote{\url{https://www.cisco.com/c/en/us/solutions/collateral/executive-perspectives/annual-internet-report/white-paper-c11-741490.html}} outlined how DDoS attacks have increased over the years (2018-2023). Figure \ref{DDoS Attacks Over the Years} shows the number of DDoS attacks during the same period. In 2023 alone, a total of 15.4 million DDoS attacks were recorded.  
According to Kaspersky's press release\footnote{\url{https://www.kaspersky.com/about/press-releases/2023_kaspersky-reports-growth-in-gamer-cyberattacks-in-2023}}, the gaming and gambling industry faced the highest number of DDoS threats between 2022-2023. This is because nearly 40\% of the world's population forms the gaming community, which is why it becomes an attractive target for the attackers. To make things more interesting, it is found that Fridays recorded the busiest day of attacks with approximately 15.36\%, while the lowest are recorded on Thursdays with 12.99\% \footnote{\url{https://www.getastra.com/blog/security-audit/ddos-attack-statistics/}}. \\

From a security perspective, the last few years have been very eventful with several high-profile data breaches, threats, and attacks against web applications\footnote{\url{https://www.enisa.europa.eu/publications/enisa-threat-landscape-2023}}. Web application attacks are possible due to existing vulnerabilities that put both end users and businesses at risk\cite{taguinod2015toward}. An effective defense approach is essential to detect and secure a web application in a timely way. Typically, a defense system may consist of four modules- i) Detection module which tries to analyze data for specific attack occurrence, ii) A prevention module which tries to prevent to attack from occurring (may be at source end or the victim end), iii) A mitigation module which tries to employ mitigation techniques (such as blocking some IPs) after attack occurrence, and iv) A tolerance module which tries to provide services to the legitimate users even when an attack is occurring. In addition, a defense approach may follow a centralized approach or even a distributed approach. When developing a defense approach for the detection of attacks in web applications, the following important points should be noted.
\begin{enumerate}
    \item Most web applications are in themselves heavily complicated because of the technologies used, hence the design of the defense approach should be such that it does not further add to the application's complexity. 
	\item The operations of legitimate users should not be harmed by the deployment of a defense approach.
	\item A defense approach should be scalable and robust.
\end{enumerate}
\subsection{Motivation}
In terms of sample sizes and dimensions, the amount of data that is currently available has increased significantly. Despite the fact that a lot of data is produced, not all of it is of high quality to be used in predictive data analysis \cite{jain2020overview}. At the same time, data may be continuously arriving at regular intervals. For such cases, the learning models need to be trained from scratch which is again computationally expensive and inefficient\cite{giraud2000note}. To offer valuable insights to predictive modeling, machine learning algorithms need relevant, independent, intelligible, meaningful, and recent data. In light of this, feature selection is essentially an important pre-processing step\cite{dhal2022comprehensive}. It aids a learning model in simplifying the learning process so that it can acquire essential and vital knowledge for prediction tasks. All the mentioned reasons have collectively motivated for the development of an incremental feature selection technique which focuses on selecting relevant and irredundant (independent) features to achieve best predictive performance. The incremental nature of the method helps to avoid processing the whole data from scratch when new data instances arrive at regular intervals.

\subsection{Contribution}
The primary contribution of this paper is a detection method for HTTP flooding attacks in the application layer. The detection method is based on an incremental feature selection method called INFS-MICC. It is incremental in the sense that it can handle incoming data incrementally and can identify a subset of highly relevant and irredundant feature subset without processing the whole data from scratch. This process is beneficial because, in a way it has the ability to memorize thereby saving valuable time and computational resources. Table \ref{Symbol Table for the Proposed Method (INFS-MICC)} gives the symbols used to describe the proposed method. The effectiveness of the proposed method is established in terms of high accuracy for three well-known HTTP-Flooding datasets.
\begin{table}[]
	\centering
	\caption{Symbol Table for the Proposed Method (INFS-MICC)}
	\label{Symbol Table for the Proposed Method (INFS-MICC)}
	\resizebox{0.5\columnwidth}{!}{%
		\begin{tabular}{@{}llll@{}}
			\toprule
			Symbol & Symbol Meaning & Symbol & Symbol Meaning \\ \midrule
			D & Dataset & $F^{old}_2$ & \begin{tabular}[c]{@{}l@{}}Contains highly relevant\\ features from $F'_{D_{old}}$\end{tabular} \\
			s & No. of samples in D & $F^{new}_3$ & \begin{tabular}[c]{@{}l@{}}Contains highly relevant\\ features from $F'_{D_{new}}$\end{tabular} \\
			d & Dimension of the dataset D &  \\ 
			$T_{m}$ & Time &  $F^{D}$ & \begin{tabular}[c]{@{}l@{}}Combination of  \\ $F^{common}_1$, $F^{old}_2$ and $F^{new}_3$ \end{tabular}    \\           
			$T_{n}$ & Time & I & Mutual Information \\
			$D_{old}$ & Data arriving at time $T_{m}$ & M & Random variable \\
			$D_{new}$ & Data arriving at time $T_{n}$ & N & Random variable \\
			F & Original feature set of D & m & \begin{tabular}[c]{@{}l@{}}Marginal probability\\ distribution of M\end{tabular} \\
			$F'_{D_{old}}$ & \begin{tabular}[c]{@{}l@{}}Feature subset containing\\ relevant and irredundant \\ features from $D_{old}$\end{tabular} & n & \begin{tabular}[c]{@{}l@{}}Marginal probability\\ distribution of N\end{tabular} \\
			$F'_{D_{new}}$ & \begin{tabular}[c]{@{}l@{}}Feature subset containing\\ relevant and irredundant \\ features from $D_{new}$\end{tabular} & p(m,n) & \begin{tabular}[c]{@{}l@{}}Joint probability \\ distribution of M and N\end{tabular} \\
			$F^{common}_1$ & \begin{tabular}[c]{@{}l@{}}Contains common features \\ from $F'_{D_{old}}$ and $F'_{D_{new}}$\end{tabular} & $C_{k}$ & Target class \\ \bottomrule
		\end{tabular}%
	}
\end{table}
\section{Related Work}
Tremendous amount of research efforts have been made to detect HTTP flooding attacks at the earliest with minimum false alarms. In the literature, researchers have categorized defense mechanisms for detecting HTTP flooding attacks based on disparate ideas. 
Zargar et al. \cite{zargar2013survey} categorize the mechanisms based on the deployment site. Destination based mechanisms deploy their defenses at the victim end, i.e., at the Web server end, and Hybrid mechanisms are deployed on both the client and the server. Praseed et al. \cite{praseed2018ddos} propose a taxonomy where the detection mechanisms are classified according to request dynamics (traffic estimation, request statistics like entropy-based measures), and request semantics (request composition and request sequence). Singh and De \cite{singh2017mlp} use a multi layer perceptron (MLP) to detect HTTP flooding attacks with features such as HTTP requests count, number of the IP addresses. For learning and adjusting the weights of the perceptrons genetic algorithm is used. The main advantage of this method is that it provides a high detection accuracy and a low false positive rate. Zhao et al. \cite{zhao2018classification} try to differentiate between flash crowd and application layer DDoS attacks. Two measures based on entropy namely, EUPI (Entropy of URL per IP) and EIPU (Entropy of IP per URL) are used. The main idea behind using such measures is that normal requests vary in size, speed and intent and as such have a high entropy value. Whereas attack packets are more similar to one another and thus have low entropy value. Wen et al. \cite{wen2010cald} propose a traffic estimation based defense mechanism which focuses on request dynamics. If the request rate is above an expected threshold at a particular point of time it means something abnormal (either an attack or flash crowd) is going on. Kalman filter is used as a measure for this purpose. Additionally, source IP distribution is used to actually identify an attack flood. \\
In \cite{yatagai2007detection} Yatagai et al. proposes an HTTP-GET flood attack detection method where the underlying idea is to analyse the page access behavior based on two detection methods. First method finds the sequence in which the pages are browsed by a user and the second measures the correlation between browsing time of a page and its information size. The downside to the first method is its low detection accuracy, however it prioritizes acitivities of normal clients, meaning normal clients will not be wrongly barred from their usual activity. On the other hand, the second method promises high detection accuracy but may misclassify a normal client. 
Dhanapal and Nithyanandam \cite{dhanapal2019slow} proposes an OpenStack based testbed framework which detects HTTP-Flooding attacks in the cloud computing platform. According to the authors detection of such attacks in the cloud is difficult because of the existence of numerous potential attack paths. Their method is highly accurate in detecting low-rate attacks in the early stages. Similar, techniques for protecting cloud computing platforms are also proposed in \cite{dhanapal2021openstack} and \cite{albaroodi2015proposed}. Mohammadi et al. \cite{mohammadi2023httpscout} propose HTTPScout, a security module which helps detect and mitigate flooding attacks using machine learning and Software Defined Networks (SDN). The proposed module continuously observes the incoming HTTP traffic flows. If any particular flow is sensed to be malicious, its source is blocked at the edge switch. This way the valuable network resources are safeguarded from the adversaries. A similar detection method is also proposed in \cite{yungaicela2021sdn}, where the authors consider both transport layer and application layer attacks in a modular SDN-based architecture. For detection both machine learning and deep learning models are employed. On the other hand, the Mininet emulator\footnote{\url{http://mininet.org/}} and Open Network Operating System\footnote{\url{https://opennetworking.org/onos/}} (ONOS) SDN controller is also deployed for implementation in a simulated environment. In the recent years, several such methods to detect application layer DDoS attacks in Software Defined Networks (SDN) have gained popularity \cite{zainudin2022efficient},\cite{yungaicela2022flexible}\cite{yousuf2022ddos}. 
The most typical presumption according to many is monotonicity, which is of the notion that adding more features would be beneficial for a learning system perform better \cite{john1994irrelevant}\cite{langley1994selection}. Researchers in machine learning and knowledge discovery who are interested in enhancing algorithm performance have over the years given feature selection a lot of interest. 
Since many learning algorithms may fail or take an excessive amount of time to run before data is reduced, feature selection is a very crucial step in pre-processing when dealing with enormous data. 
Feature selection methods in the literature are mainly categorized into: Filter, Wrapper and Embedded methods. \\

In order to give an ordered list of feature ranks, filter methods use statistical measures such as information gain, correlation, and mutual information \cite{guyon2003introduction}\cite{lazar2012survey}\cite{jovic2015review}. These ranking systems aid in highlighting the characteristics that are crucial. Prior to executing the classification task, irrelevant features are filtered out and eliminated because their presence does not help improve the performance of a machine learning algorithm. To maximize the advantages of competitive ranking, numerous filter methods have been utilized in conjunction with population-based heuristic search methodologies \cite{usman2020filter} \cite{zhu2017improved}\cite{hoque2014mifs} \cite{binder2020multi}. Feature-feature and feature-class mutual information are used in the widely used MIFS (Mutual Information Feature Selection) method to choose a feature subset that maximizes classification accuracy \cite{battiti1994using}.\\
Two categories of wrapper feature subset selection methods that are often used in the literature are sequential selection algorithms and heuristic search algorithms \cite{el2016review}\cite{jovic2015review}\cite{chandrashekar2014survey}\cite{miao2016survey}\cite{kohavi1997wrappers}. To choose feature subset, Maldonado and Weber \cite{maldonado2009wrapper} suggest a sequential backward selection wrapper approach utilizing Support Vector Machines (SVMs). Using cosine similarity and SVMs, Gang and Jin \cite{chen2015novel} choose relevant and independent features. On the other hand, Hsu et al. \cite{hsu2011hybrid} provide a hybrid feature selection method, where the filter methods assist in effectively finding the candidate features and the wrappers are in charge of delivering the subset of features that ensures the best possible classification accuracy. In order to produce distinct lists of ranked features, ensemble feature selection methods like \cite{sarmah2022cost} rely on base feature selection algorithms. The final ranked list of features is created by combining these individually ranked features based on a score. In conjunction with four search algorithms, including ensemble forward and backward sequential selection, hill-climbing, and genetic search, Tsymbal et al. \cite{tsymbal2005diversity} examine diversity metrics to assess diversity in the ensemble feature selection methods.

\section{Background}
This section presents the background of our method. It exploits the power of mutual information and correlation measures to design the proposed feature selection method. 
\subsection{Mutual Information for Feature Selection} 
Mutual Information is important for feature selection since it helps determine how pertinently a specific feature (or characteristic) is related to the target class. In other words, it enables the assessment of a feature's predictive value for a given class. A feature, say $f_i$, gives more information on the target if it has a higher mutual information score with the target class compared to another feature, say $f_j$. So, $f_i$ is more valuable in predicting the target class.\\
Let's suppose that there are two random variables named \textit{M} and \textit{N}. \textit{Mutual Information} is the amount of information that \textit{$M$} knows about \textit{$N$}. This can be expressed mathematically as in Equation \ref{mutual_info_equation1}, where \textit{m} and \textit{n} are the marginal probability distributions for \textit{M} and \textit{N} respectively. 
\begin{equation}
	\label{mutual_info_equation1}
	I (M;N) = \sum_{m,n} p(m,n) \log \frac{p(m,n)}{p(m)p(n)}
\end{equation}
The joint probability distribution function for the random variables \textit{M} and \textit{N} is actually expressed as \textit{$p(m,n)$} in Equation \ref{mutual_info_equation1}, while \textit{$p(m)$} and \textit{$p(n)$} denote the marginal probability distributions for \textit{M} and \textit{N}. It is important to note that the Mutual Information between \textit{M} and \textit{N} is said to be zero if they are statistically independent.

\subsection{Correlation for Feature Selection}
In feature selection, to determine the relationship between two features, such as $f_i$ and $f_j$, the statistical measure of correlation is used. Two attributes can have a positive, negative, or zero correlation value depending on how they are related to each other. If they are closely linked, including just one of them in the feature set is sufficient. On the other side, having both features would make the feature set superfluous. Therefore, the goal is to select a subset of features from the original feature set with the least amount of overlap between them.\\
The linear relationship between two entities, let's say \textit{M} and \textit{N}, is described by Pearson's correlation coefficient as shown in Equation \ref{pearson_equation1}. The value of the correlation coefficient ranges from -1 to +1. A high negative correlation is represented by -1, and a high positive correlation is represented +1. Additionally, a value of 0 denotes a lack of association between the two entities. For the proposed method, the absolute value of the correlation coefficient is taken into account because the relationship's strength is of interest and not its direction of positive or negative. 
\begin{equation}\label{pearson_equation1}
	Corr - Coeff = \frac{\sum {(m_i - \bar{m})(n_i - \bar{n})}}{\sqrt{\sum{{{(m_i - \bar{n})}^2 \sum{{(n_i - \bar{n})}^2}}}}}
\end{equation}
The aim is to identify an optimal subset of features that are both highly relevant and non-redundant. Mutual information is used to ensure relevance, while Pearson's correlation measure ensures irredundancy.

\section{Problem Definition}
Let's assume a dataset D containing $s$ samples. $D_{old}$ comprises of samples that arrived at time $T_m$ and $D_{new}$ is the incremental batch of data newly arrived at time $T_n$ ($T_m < T_n$). Dataset D is characterized by the feature set $F=\{f_1, f_2, f_3, ....f_d\}$, where d is the dimension of the dataset. Next task is to find, $F'_{D_{old}}$ which is the feature subset containing relevant (high feature-class mutual information) and irredundant (low feature-feature correlation) features selected from $D_{old}$. Now, for the given data increment $D_{new}$, the problem is to identify the optimal subset of features, $F^D$ for the whole dataset i.e. $D_{old} \cup D_{new}$ i.e D, ensuring best possible accuracy and without starting the computation from scratch.
\section{Proposed Work}
\begin{figure}[h]
	\centering
	\includegraphics[width=0.85\textwidth]{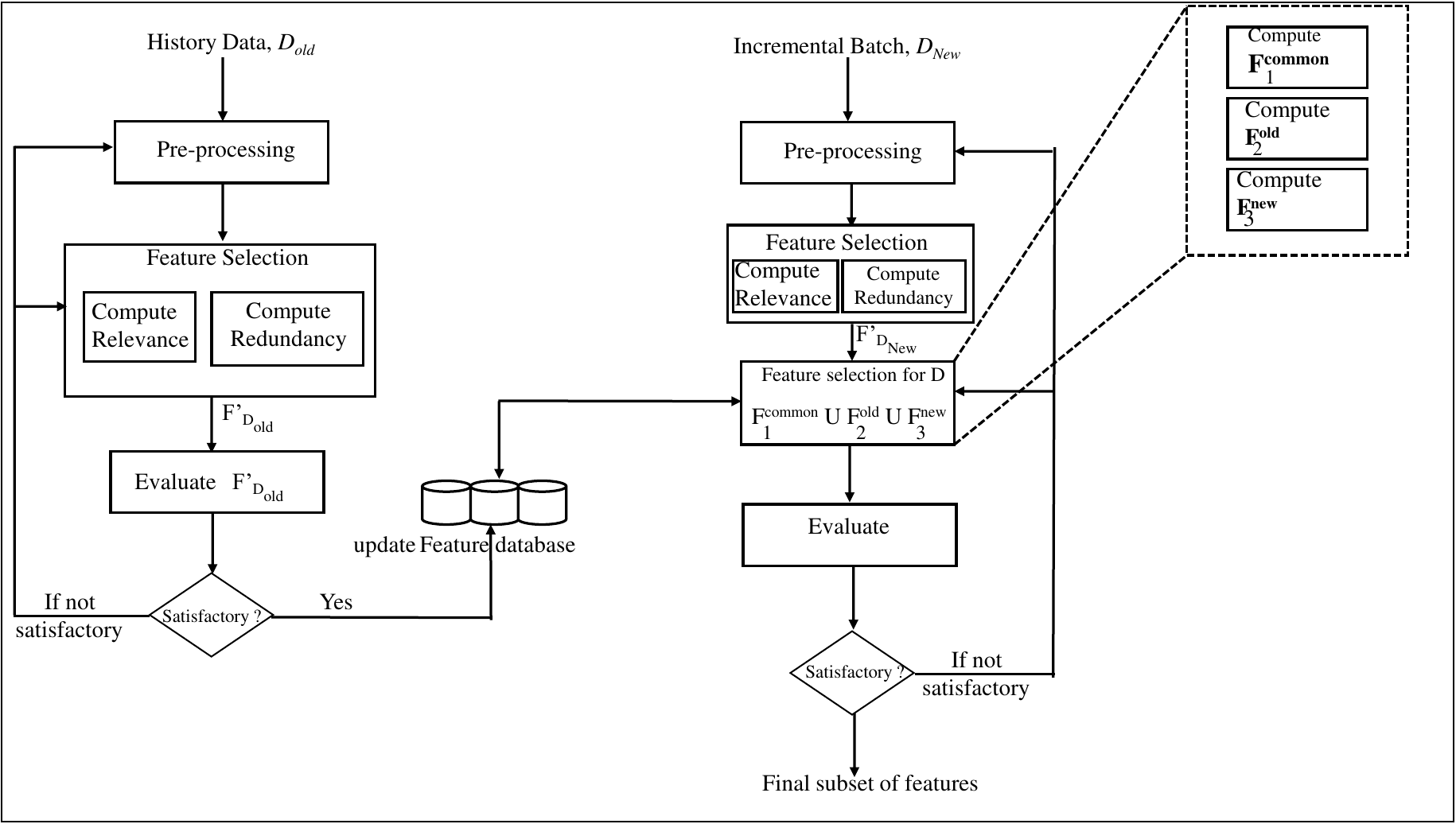}
	\caption{Proposed Framework for INFS-MICC}
	\label{Proposed Framework for INFS-MICC}
\end{figure}

The proposed method, INFS-MICC selects a ranked optimal subset of features to detect HTTP-Flooding attacks. 
The main attraction of our method is its ability to handle incremental data while selecting the subset of features to ensure best possible classification accuracy. 
It avoids re-doing the entire computation from scratch to gain new knowledge. It makes use of the previous results obtained from the original data along with new results from the added-in data. The proposed method is depicted in Figure \ref{Proposed Framework for INFS-MICC}.\\ 
The feature sets $F^{common}_1, F^{old}_2, F^{new}_3,$ and $F^{D}$ described in Figure \ref{Proposed Framework for INFS-MICC} are explained mathematically in Equation \ref{fcommon}, \ref{fold}, \ref{fnew} and \ref{fD}.  \\
\begin{equation}\label{fcommon}
	F^{common}_1 = F'_{D_{new}} \cap F'_{D_{old}}
\end{equation}
\begin{equation}\label{fold}
	F^{old}_2 = \{f_i \in F'_{D_{old}} | rank(f_i) \ge \alpha \}
\end{equation}
\begin{equation}\label{fnew}
	F^{new}_3 = \{f_j \in F'_{D_{new}} | rank(f_j) \ge \alpha\} 
\end{equation}
\begin{equation}\label{fD}
	F^{D} = F^{common}_1 \cup F^{old}_2 \cup F^{new}_3
\end{equation}
Primarily, three tasks are performed. First, the missing values if any are estimated, by averaging the column values. Second, features which have zero variance are eliminated because they do not contribute in the decision-making during prediction. Third, to bring the values to a uniform range of 0 to 1, min-max normalization technique is used. \\
Next, the proposed feature selection method is applied which is designed to handle incremental data or added-in data. The feature selection process is mainly based on computing the relevance of the features in terms of mutual information and computing the redundacny among the features in terms of correlation. 
\subsection{Specifics of INFS-MICC}
To assess the strength of the relationship between features correlation is used, specifically Pearson's correlation coefficient. Higher strength indicates that the features are highly correlated. The proposed method uses Pearson's correlation coefficient because of three main reasons: $ i)$ It measures the relationship between two continuous variables (raw values of the variables), unlike other correlation measures such as Spearman Rank Correlation which takes into consideration the ranks of the data or Kendall's Tau correlation measure which considers the ordinal association between two variables \cite{rickert2023efficiency}, $ii)$ It is a widely accepted standard measure and hence is not influenced by the scales of the continuous valued features \cite{kim2018statistical}, $iii)$ Pearson's correlation coefficient is simple and fast in terms of computational complexity ($\mathcal{O}(n)$) compared to Spearman's coefficient ($\mathcal{O}(n\log{}n)$) \cite{jaskowiak2010comparative}. Additionally, as already mentioned correlation measures (in this case Pearson's correlation measure) are sensitive to outliers. This drawback is overcome with a two fold solution. First, mutual information (as it is insensitive to outliers \cite{correa2013mutual}) is introduced to the proposed score. Second, to negate out the outlier effects of a feature say $f_j$ when calculating the correlation with feature $f_i$, the average correlation of $f_i$ is introduced and subtraction of the two entities is performed as shown in Equation \ref{micc_ud}. This ensures that the outlier values of $f_j$ will not influence the values of feature $f_i$. \\
\subsection{Relevance and Independent Feature Subset Finding}
INFS-MICC is an incremental feature selection method, which assumes that data arrives in batches. For simplicity, two batches of data $D_{old}$ (already arrived) and $D_{new}$ (now arrives) are considered. First, it is assumed that for $D_{old}$, the feature subset $F^{'}_{D_{old}}$ is selected using the score described in Equation \ref{micc_ud}.
\begin{equation}\label{micc_ud}
	MICC\mbox{-}UD
	{(f_i)} = \frac{Relevance\_{score}(f_i,C)}{max_{i \neq j}(|{avg\_Corr}|(f_i) - Corr(f_i,f_j))}
\end{equation}
The relevance score and average correlation (avg\_Corr) mentioned in Equation \ref{micc_ud} is calculated as shown below in Equation \ref{relevance_score} and Equation \ref{average_correlation}. 
\begin{equation}\label{relevance_score}
	Relevance\_{score}(f_i,C) = Mutual Information (f_i,C)
\end{equation}
\begin{equation}\label{average_correlation}
	avg\_Corr(f_i) = \frac{\sum_{\substack{j=1 \\ i \neq j}}^{d}(Corr(f_i,f_j))}{d} 
\end{equation}
Over time, as incremental batch of data $D_{new}$ becomes available, the respective feature subset $F^{'}_{D_{new}}$ is also identified using the same approach. It is important here to note that, data batch $D_{new}$ may acquire some new features over time. So, $F^{'}_{D_{new}}$ may contain some features which were not previously present in $F^{'}_{D_{old}}$ and it may or may not be relevant for $D_{old}$. To address this issue, the feature subset $F^D$ is calculated. $F^D$ is a combination of three feature subsets namely $F^{common}_1$, $F^{old}_2$ and $F^{new}_2$ as shown in Equation \ref{fD}. $F^{common}_1$ contains the common features from $F^{'}_{D_{new}}$ and $F^{'}_{D_{old}}$ as shown in Equation \ref{fcommon}. $F^{old}_2$ contains the highly relevant (i.e. features with $rank \ge \alpha$, a user defined threshold) features from $F^{'}_{D_{old}}$ as described in Equation \ref{fold}. Similarly, $F^{new}_3$ contains the highly relevant features from $F^{'}_{D_{new}}$ as shown in Equation \ref{fnew}. The feature set $F^D$ is now used to evaluate the new batch data, $D_{new}$. If performance is found to be satisfactory, then a complete scan and evaluation of $D_{old}$ can be avoided. 
\subsection{Optimal Feature Subset Identification using Recursive Feature Elimination}
	After obtaining the ranked list of features, next the optimal feature subset needs to be identified. Optimal subset of features mean adding features to this subset does not increase the classifier performance and at the same time, removing any feature from the subset deteriorates the classifier performance. Here, the optimal feature subset is obtained by recursively eliminating the features from the ranked list. The recursive feature elimination step works primarily with five predictors for making the final predictions.\\ 
Following definitions provide the theoretical basis of the proposed method. \\
\begin{definition}{(Feature Relevance)} \label{relevant feature}
	For any feature $f_i$, its relevance is defined in terms of mutual information to the target class $C_k$. Higher the mutual information score for $f_i$, higher is its relevance to $C_k$. 
\end{definition}
\begin{definition}{(Highly Relevant)}\label{highly_relevant}
	A feature $f_i$ is said to be highly relevant iff any of the following two cases holds.\\
	\textit{Case 1}: $f_i$ $\in$ $F^{'}_{D_{new}}$, and relevance($f_i$,$C_k$) $> \alpha$ in $D_{new}$ and $f_i$ $\notin$ $F^{'}_{D_{old}}$, \\ where 0 $<$ $\alpha$ $<$ 1, a user defined threshold.\\
	\textit{Case 2}: $f_i$ $\in$ $F^{'}_{D_{old}}$ , and relevance($f_i$,$C_k$) $> \alpha$ in $D_{old}$ and $f_i$ $\notin$ $F^{'}_{D_{new}}$, \\ where $0 < \alpha < 1$, is a user defined threshold. 
\end{definition}
\begin{definition}{(Independence of a Feature)}
	The independence of a feature $f_i$ is defined in terms of average correlation with respect to all other features in feature set F. Feature $f_i$ has high independence if its average correlation score with all other features is low. 
\end{definition}
\begin{proposition}\label{Proposition 11}
    A feature $f_i \in F^D$ is relevant and irredundant if and only if  $f_i \in F'_{D_{new}}$ or $f_i \in F'_{D_{old}}$. 
	\begin{proof}
		Suppose $f_i \in F^D$, however, $f_i \notin F'_{D_{new}}$ or $f_i \notin F'_{D_{old}}$. A feature $f_i$ is included in $F^D$ only when it is highly relevant (Definition \ref{highly_relevant}) and independent of other features. Now, a feature $f_i$ can be highly relevant for D ($D_{new} \cup D_{old}$) only when -
		\begin{itemize}
			\item $f_i \in F'_{D_{new}}$ and $f_i \in F'_{D_{old}}$, or,
			\item $f_i$ is highly relevant for either $D_{new}$ or $D_{old}$.
		\end{itemize}
		Hence, $f_i$ must be relevant and irredundant. 
	\end{proof}
\end{proposition}
	

\begin{proposition}\label{Proposition 21}
	The feature subset selected by INFS-MICC is optimal.
	\begin{proof}
		Assume that feature subset $F^D$ selected by INFS-MICC is not optimal. In other words, there is possibility of inclusion or exclusion of feature(s) in $F^D$. However, a feature $f_i$ is included in $F^D$ only when any of the following condition is true:\\
		\emph{Condition 1:} \label{point1} $f_i \in F'_{D_{old}}$ and $f_i \in F'_{D_{New}}$ \\
		\emph{Condition 2:} \label{point2} $f_i$ is assigned higher rank for either $D_{new}$ or $D_{old}$.\\
		Further, the feature subset created considering the conditions 1 and 2 undergoes a recursive feature elimination process to find the optimal subset of selected features so that any exclusion or inclusion of feature(s) leads to deterioration of performance. Hence, the assumption does not hold and hence the proof.  
	\end{proof}
\end{proposition}

\section{Experiments and Results}
\begin{figure}[ht!]
	\centering
	\begin{minipage}{0.45\textwidth}
	\centering
	\includegraphics[width=0.9\linewidth]{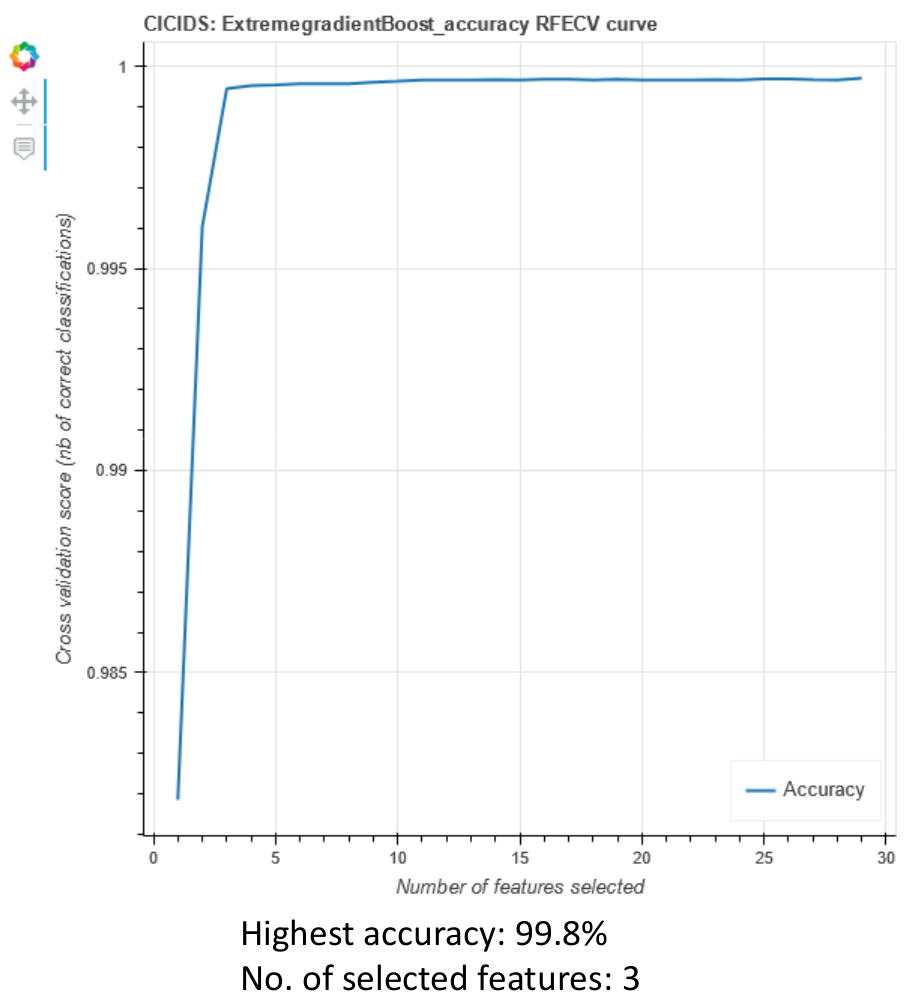}
	\caption{RFE with XGBoost Classifier for CICIDS Dataset (Accuracy)}
	\label{RFE with XGBoost Classifier for CICIDS Dataset (Accuracy)}
	\end{minipage}\hfill
	\begin{minipage}{0.45\textwidth}
		\centering
    \includegraphics[width=0.9\textwidth]{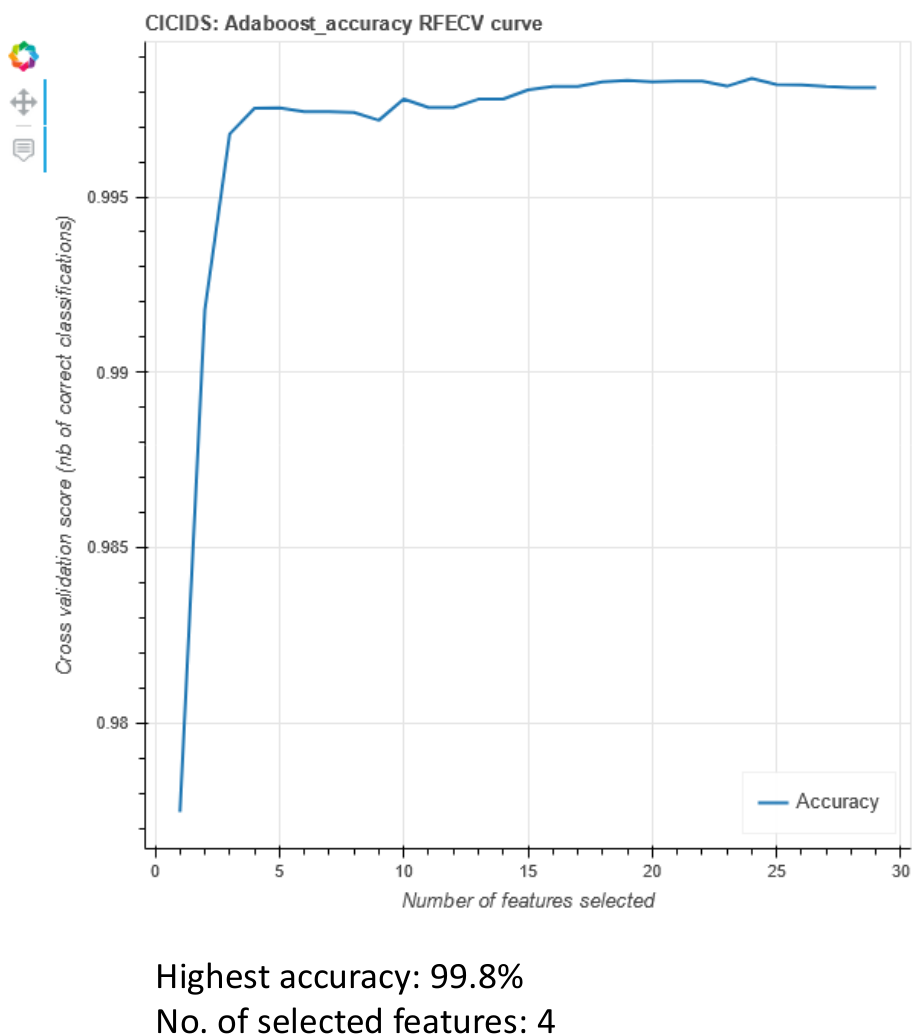} 
		\caption{RFE with Adaboost Classifier for CICIDS Dataset (Accuracy)}
		\label{RFE with Adaboost Classifier for CICIDS Dataset (Accuracy)}
	\end{minipage}

\end{figure}

\begin{figure}[ht!]
	\centering
	\begin{minipage}{0.45\textwidth}
		\centering
		\includegraphics[width=0.9\textwidth]{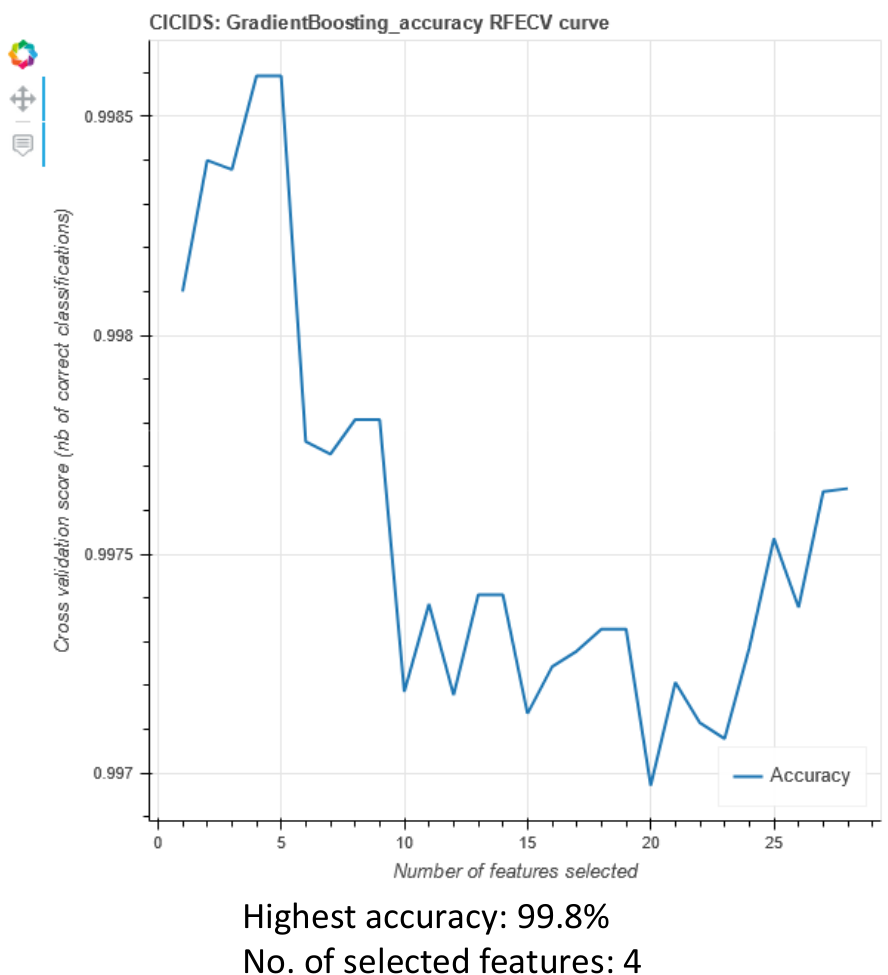} 
		\caption{RFE with Gradient Boosting Classifier for CICIDS Dataset (Accuracy)}
		\label{RFE with Gradient Boosting Classifier for CICIDS Dataset (Accuracy)}
	\end{minipage}\hfill
	\begin{minipage}{0.45\textwidth}
		\centering
		\includegraphics[width=0.9\textwidth]{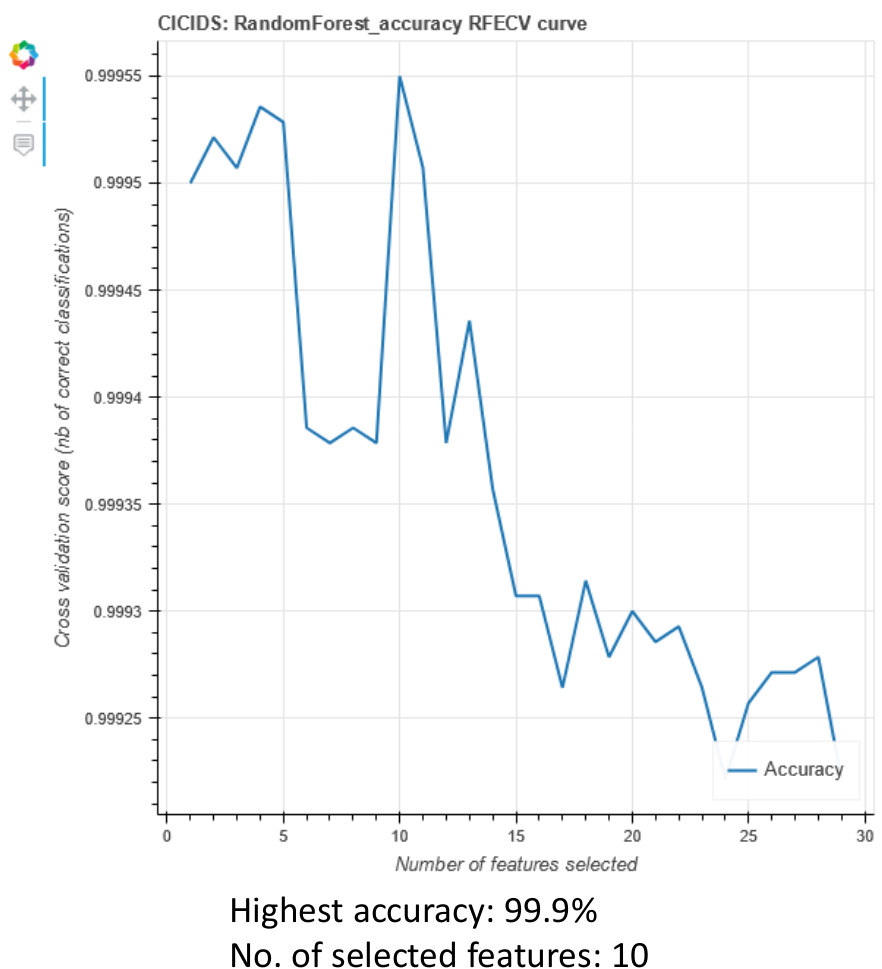} 
		\caption{RFE with Random Forest Classifier for CICIDS Dataset (Accuracy)}
		\label{RFE with Random Forest Classifier for CICIDS Dataset (Accuracy)}
	\end{minipage}
\end{figure}

\begin{figure}[ht!]
	\centering
	\begin{minipage}{0.45\textwidth}
		\centering
		\includegraphics[width=0.9\textwidth]{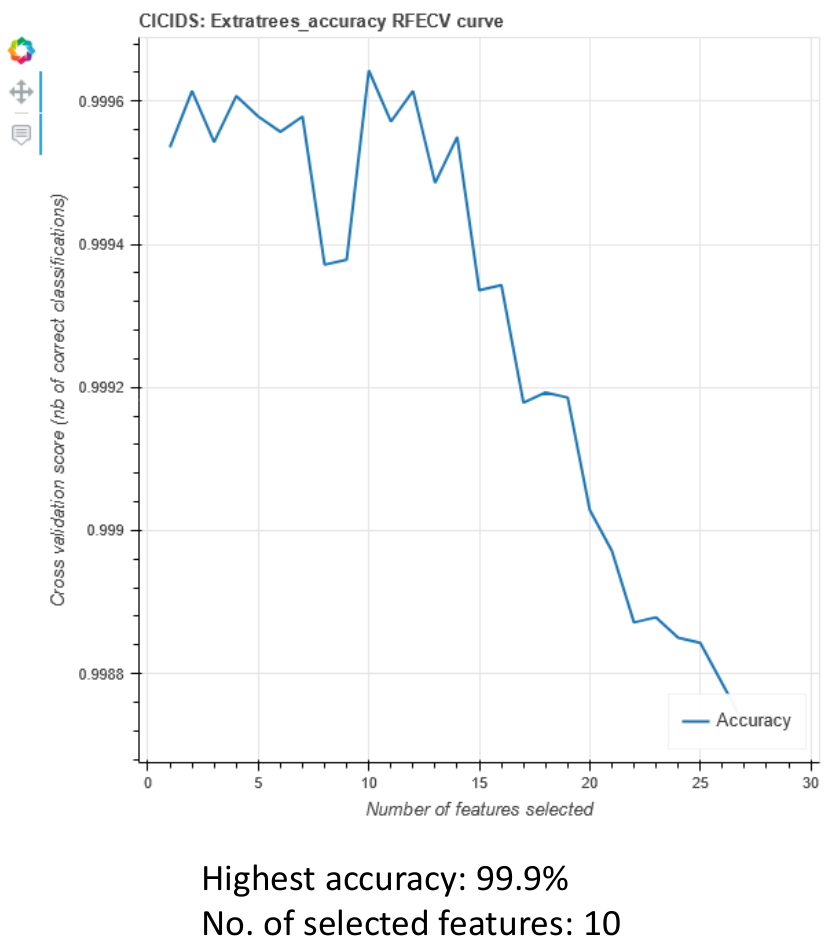} 
		\caption{RFE with Extra Trees Classifier for CICIDS Dataset (Accuracy)}
		\label{RFE with Extra Trees Classifier for CICIDS Dataset (Accuracy)}
	\end{minipage}\hfill
	\begin{minipage}{0.45\textwidth}
	\centering
	\includegraphics[width=0.9\textwidth]{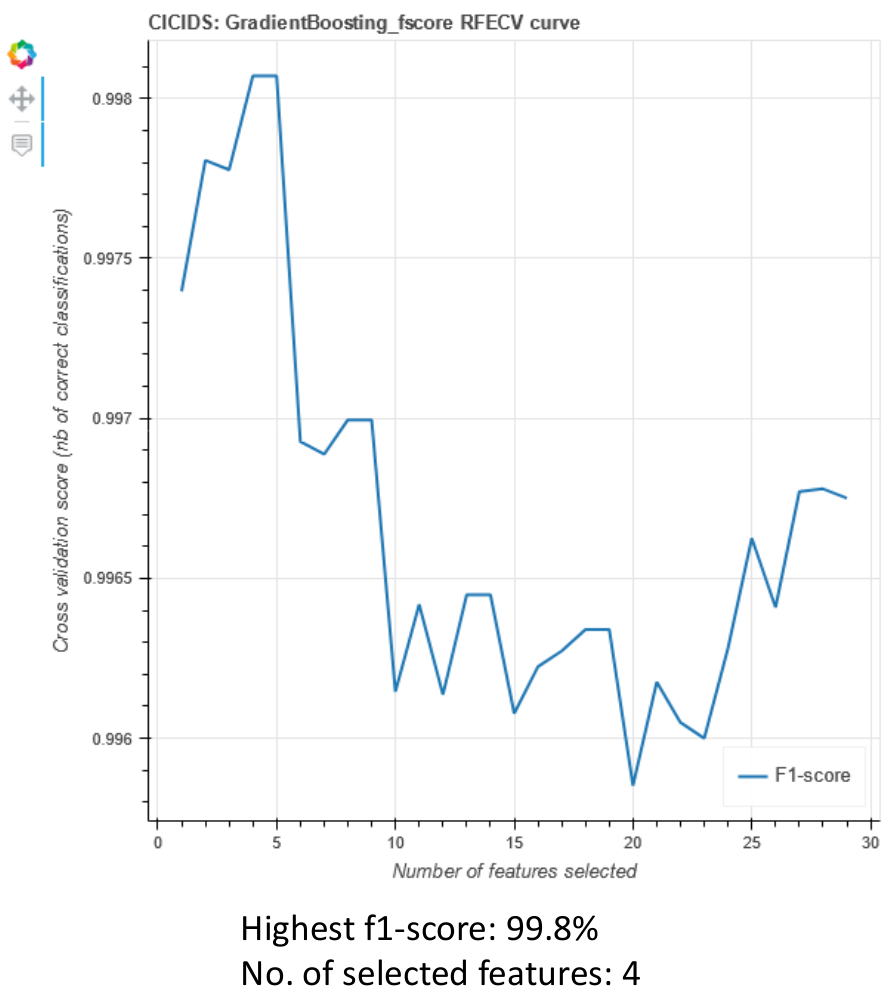} 
	\caption{RFE with Gradient Boosting Classifier for CICIDS Dataset (F1-score)}
	\label{RFE with Gradient Boosting Classifier for CICIDS Dataset (F1-score)}
	\end{minipage}
\end{figure}

\begin{figure}[ht!]
	\centering
	\begin{minipage}{0.45\textwidth}
		\centering
		\includegraphics[width=0.9\textwidth]{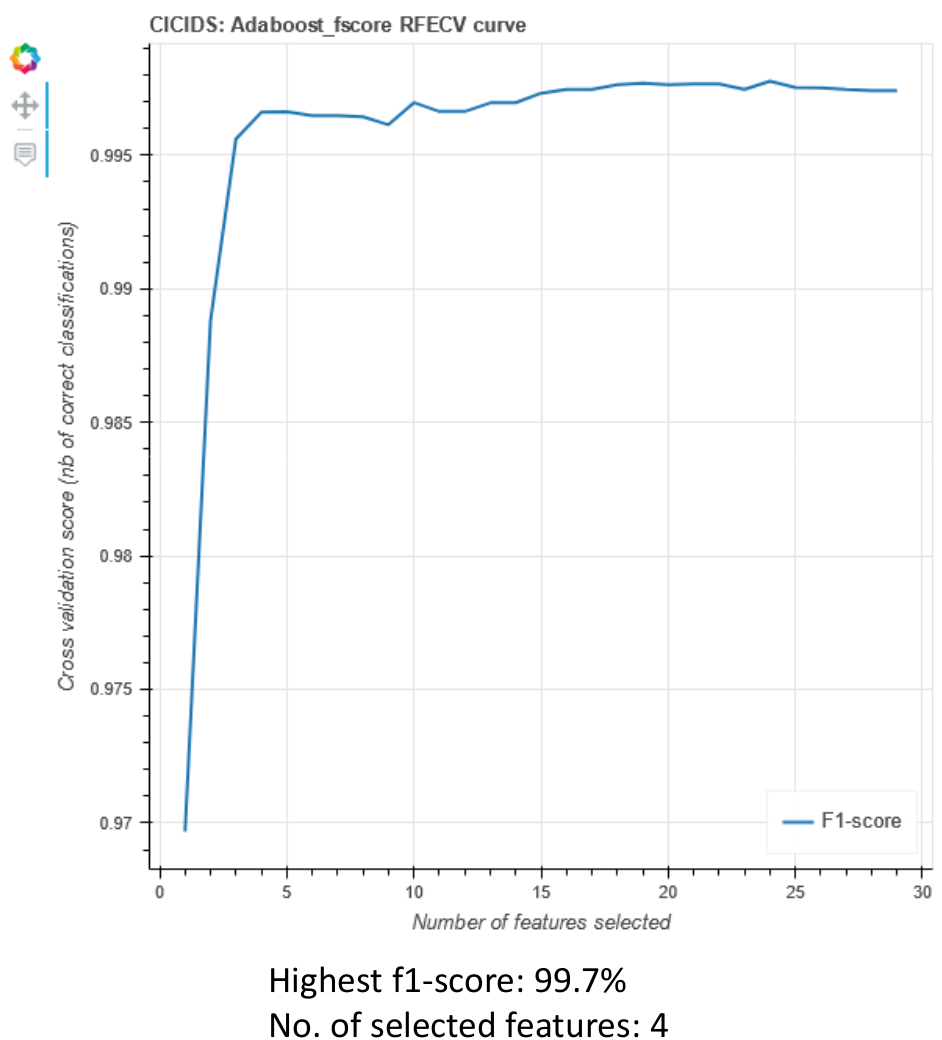} 
		\caption{RFE with Adaboost Classifier for CICIDS Dataset (F1-score)}
		\label{RFE with Adaboost Classifier for CICIDS Dataset (F1-score)}
	\end{minipage}\hfill
	\begin{minipage}{0.45\textwidth}
		\centering
		\includegraphics[width=0.9\linewidth]{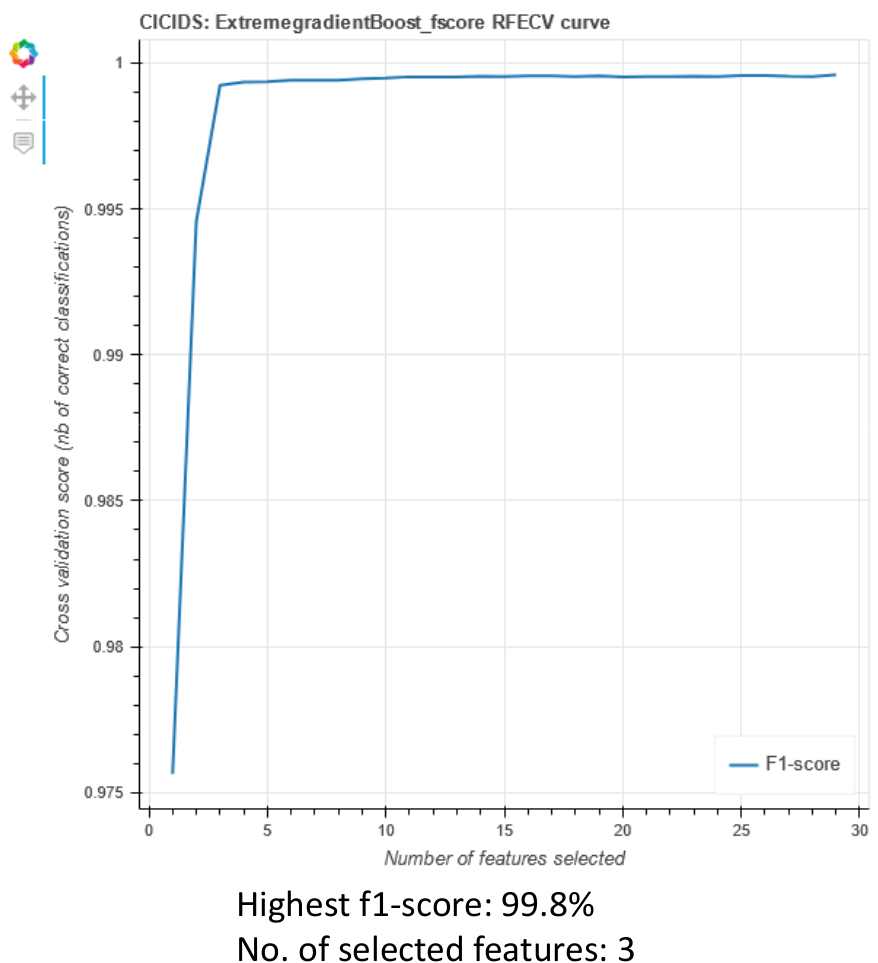}
		\caption{RFE with XGBoost Classifier for CICIDS Dataset (F1-score)}
		\label{RFE with XGBoost Classifier for CICIDS Dataset (F1-score)}
	\end{minipage}
\end{figure}

\begin{figure}[ht!]
	\centering
	\begin{minipage}{0.45\textwidth}
		\centering
		\includegraphics[width=0.9\textwidth]{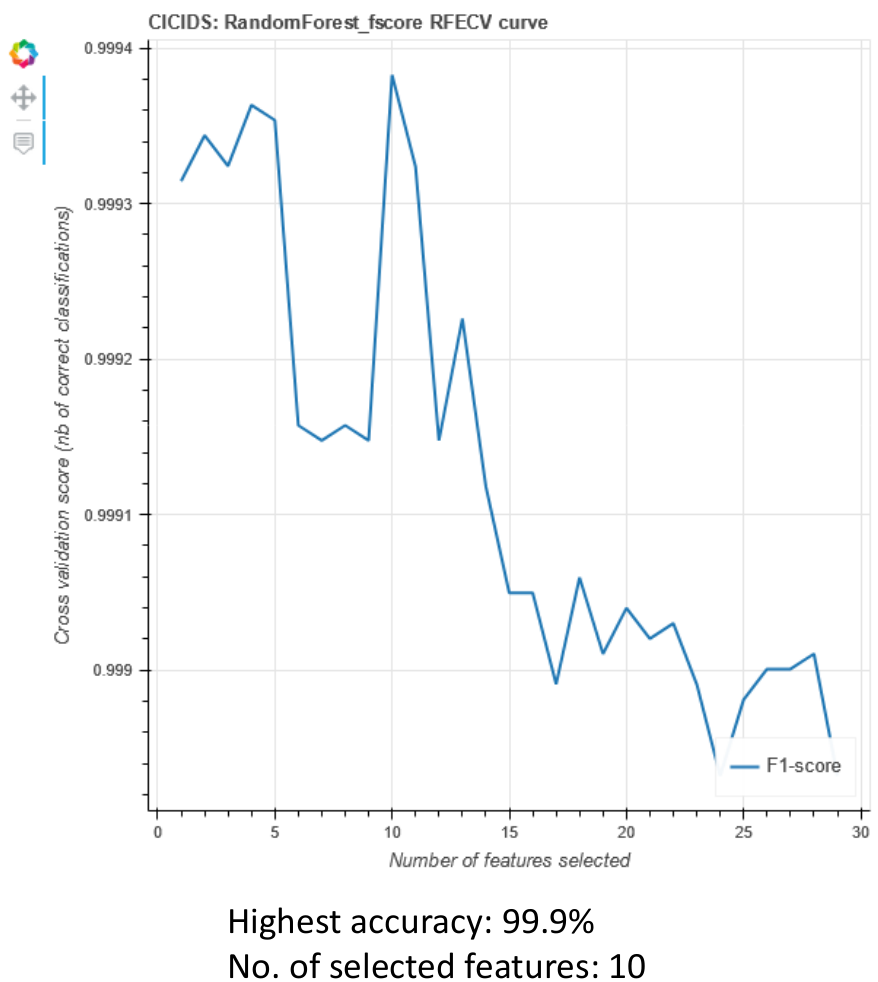} 
		\caption{RFE with Random Forest Classifier for CICIDS Dataset (F1-score)}
		\label{RFE with Random Forest Classifier for CICIDS Dataset (F1-score)}
	\end{minipage}\hfill
	\begin{minipage}{0.45\textwidth}
		\centering
		\includegraphics[width=0.9\textwidth]{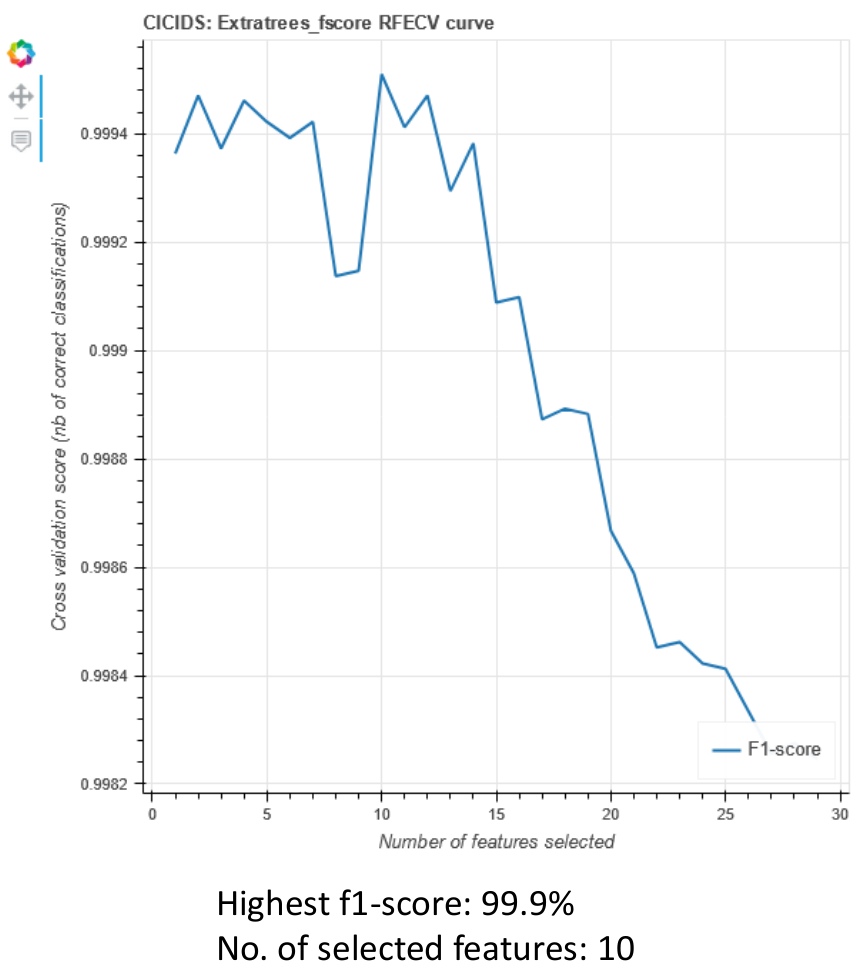} 
		\caption{RFE with Extra Trees Classifier for CICIDS Dataset (F1-score)}
		\label{RFE with Extra Trees Classifier for CICIDS Dataset (F1-score)}
	\end{minipage}
\end{figure}

This section presents the experimental results. For evaluation purposes, five different ensemble learners namely, Adaboost, Gradient Boosting, Extreme Gradient Boosting, Random Forest and Extra Trees are used. For each of these learners, the results are evaluated in terms of both Accuracy and F1-score. Along with these two measures, the number of optimal features are also presented. For this, Recursive Feature Elimination (RFE) is used in a cross validation setting. In Table \ref{Datasets Used} the datasets used for evaluating the proposed method is presented.  \\
\begin{table}[h]
	\centering
	\caption{Datasets Used}
	\label{Datasets Used}
	\resizebox{0.7\columnwidth}{!}{%
		\begin{tabular}{@{}
				>{\columncolor[HTML]{FFFFFF}}l 
				>{\columncolor[HTML]{FFFFFF}}l 
				>{\columncolor[HTML]{FFFFFF}}l 
				>{\columncolor[HTML]{FFFFFF}}l @{}}
			\toprule
			Dataset   name & No.   of instances & No.   of features & No.   of classes \\ \midrule
			HTTP Flood \tablefootnote{https://www.kaggle.com/datasets/jacobvs/ddos-attack-network-logs?resource=download} & 21,60,668 & 28 & 2 \\
			UNSW\cite{moustafa2015unsw} & 25,40,044 & 49 & 2 \\
			CICIDS\cite{jazi2017detecting} & 28,30,540 & 80 & 2 \\ \bottomrule
		\end{tabular} }
\end{table}
In case of the CICIDS dataset, the highest accuracy of 99.8\% is obtained by Extreme Gradient Boosting classifier with 3 features as shown in Figure \ref{RFE with XGBoost Classifier for CICIDS Dataset (Accuracy)}. All the other classifiers i.e AdaBoost, Gradient Boosting, Random Forest and Extra Trees show similar performance in terms of accuracy as shown in Figures \ref{RFE with Adaboost Classifier for CICIDS Dataset (Accuracy)}, \ref{RFE with Gradient Boosting Classifier for CICIDS Dataset (Accuracy)}, \ref{RFE with Random Forest Classifier for CICIDS Dataset (Accuracy)} and \ref{RFE with Extra Trees Classifier for CICIDS Dataset (Accuracy)} respectively. However, to achieve that performance the classifiers require 4 (for Adaboost), 4 (for Gradient Boosting) and 10 (for both Random forest and Extra Trees) which is higher than the number of features required by Extreme Gradient Boosting classifier. Hence, in this case, it is concluded that the optimal performance is given by Extreme Gradient Boosting classifier with 3 features. For the same dataset, F1-scores are illustrated in Figure \ref{RFE with Gradient Boosting Classifier for CICIDS Dataset (F1-score)} for Gradient Boosting, Figure \ref{RFE with Adaboost Classifier for CICIDS Dataset (F1-score)} for Adaboost, Figure \ref{RFE with XGBoost Classifier for CICIDS Dataset (F1-score)} for XGBoost, Figure \ref{RFE with Random Forest Classifier for CICIDS Dataset (F1-score)} for Random Forest, and Figure \ref{RFE with Extra Trees Classifier for CICIDS Dataset (F1-score)} for Extra Trees classifier.


%

\begin{figure}[ht!]
	\centering
	\begin{minipage}{0.45\textwidth}
	\centering
	\includegraphics[width=0.9\textwidth]{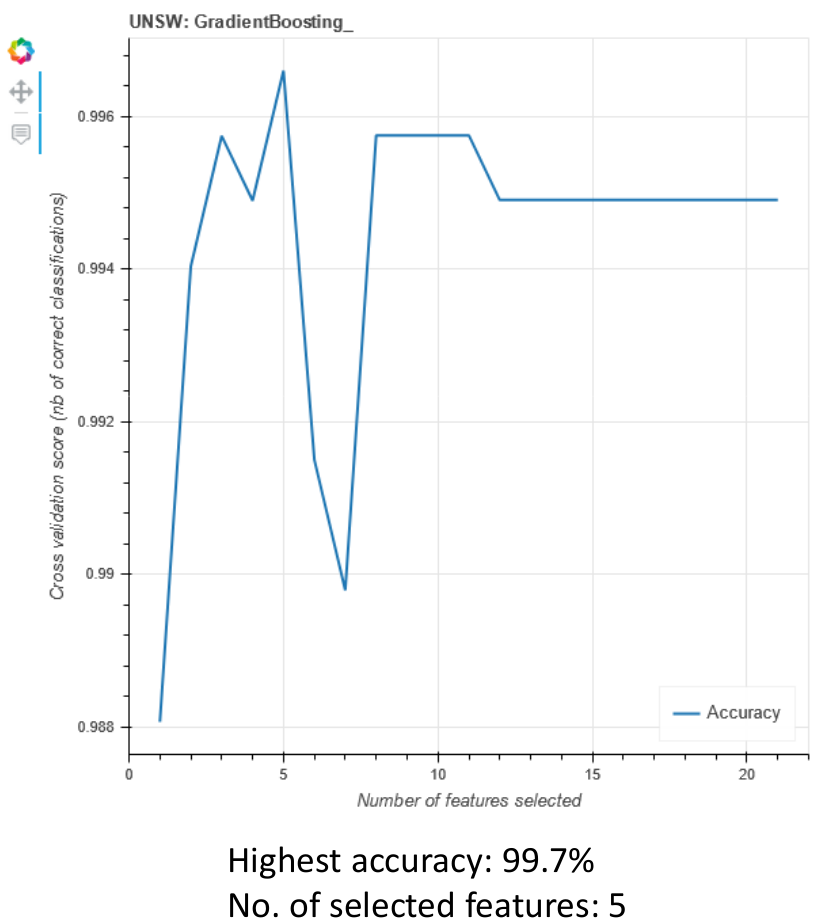} 
	\caption{RFE with Gradient Boosting Classifier for UNSW Dataset (Accuracy)}
	\label{RFE with Gradient Boosting Classifier for UNSW Dataset (Accuracy)}
	\end{minipage}\hfill
	\begin{minipage}{0.45\textwidth}
		\centering
		\includegraphics[width=0.9\textwidth]{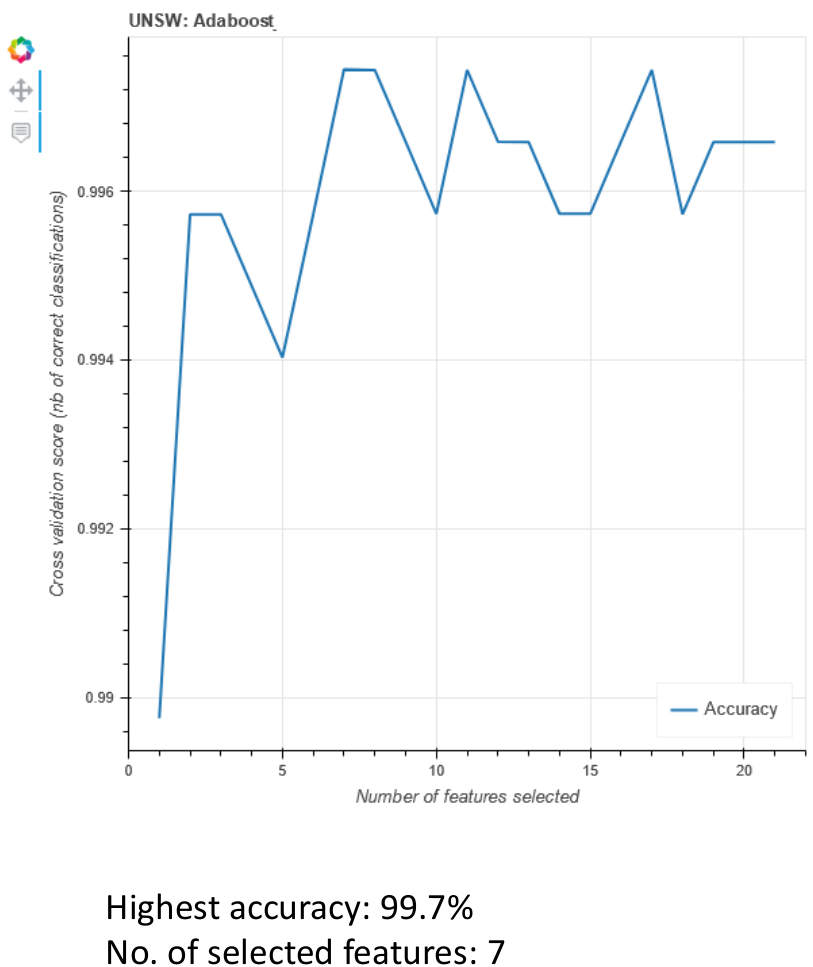} 
		\caption{RFE with Adaboost Classifier for UNSW Dataset (Accuracy)}
		\label{RFE with Adaboost Classifier for UNSW Dataset (Accuracy)}
	\end{minipage}
\end{figure}

\begin{figure}[ht!]
	\centering
	\begin{minipage}{0.45\textwidth}
		\centering
		\includegraphics[width=0.9\linewidth]{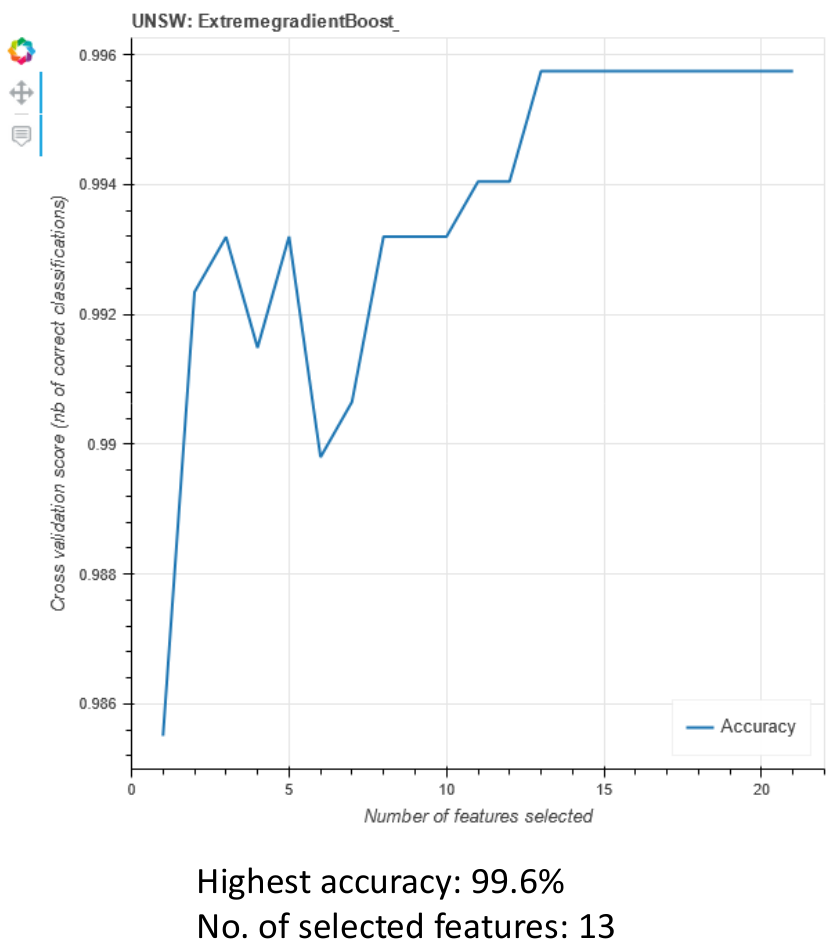}
		\caption{RFE with XGBoost Classifier for UNSW Dataset (Accuracy)}
		\label{RFE with XGBoost Classifier for UNSW Dataset (Accuracy)}
	\end{minipage}\hfill
	\begin{minipage}{0.45\textwidth}
		\centering
			\includegraphics[width=0.9\textwidth]{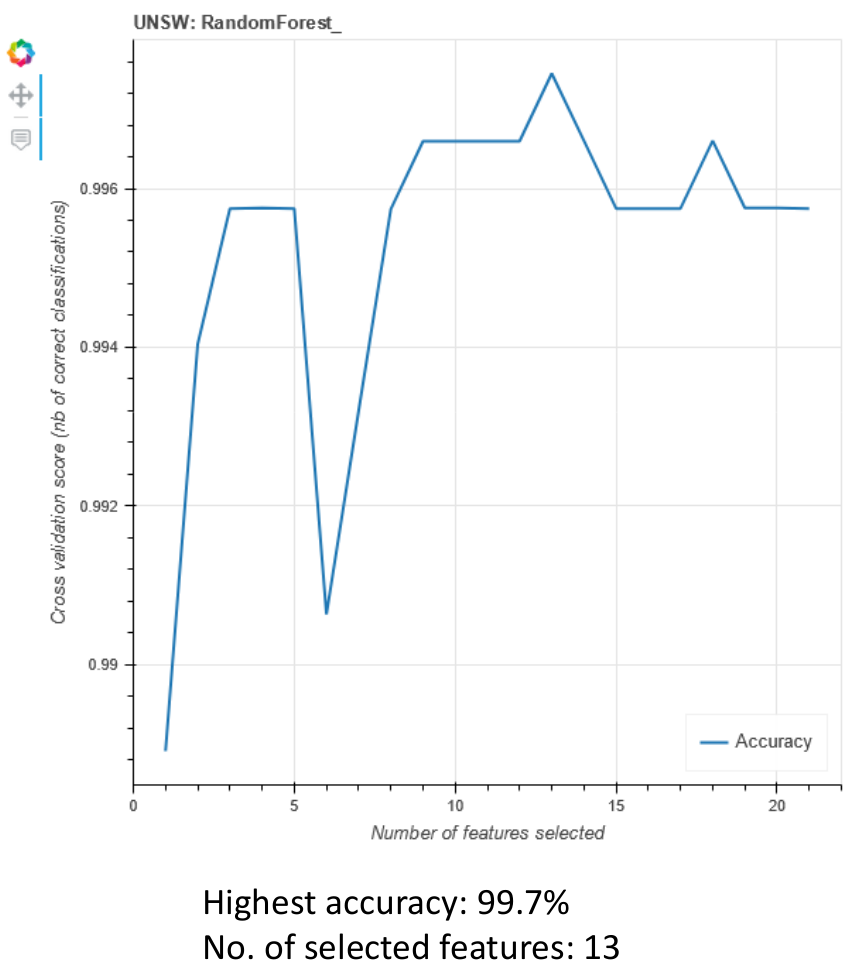} 
		\caption{RFE with Random Forest Classifier for UNSW Dataset (Accuracy)}
		\label{RFE with Random Forest Classifier for UNSW Dataset (Accuracy)}
	\end{minipage}
\end{figure}

\begin{figure}[ht!]
	\centering
	\begin{minipage}{0.45\textwidth}
		\centering
			\includegraphics[width=0.9\textwidth]{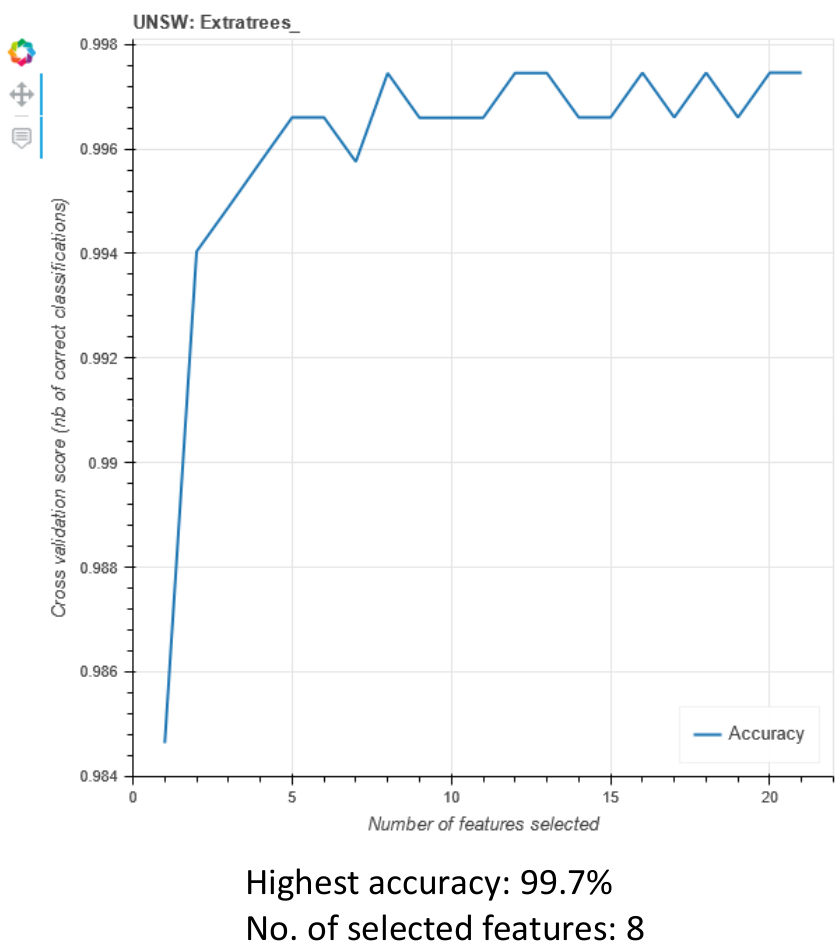} 
		\caption{RFE with Extra Trees Classifier for UNSW Dataset (Accuracy)}
		\label{RFE with Extra Trees Classifier for UNSW Dataset (Accuracy)}
	\end{minipage}\hfill
	\begin{minipage}{0.45\textwidth}
		\centering
	\includegraphics[width=0.9\textwidth]{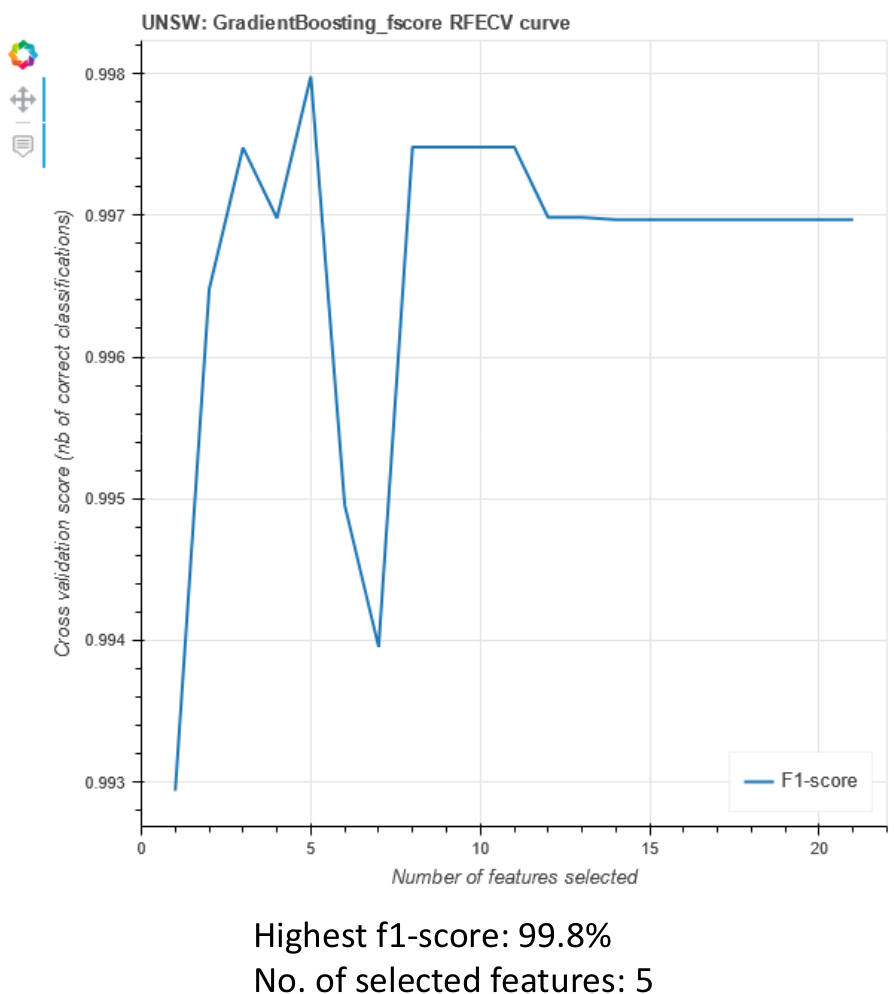} 
	\caption{RFE with Gradient Boosting Classifier for UNSW Dataset (F1-score)}
	\label{RFE with Gradient Boosting Classifier for UNSW Dataset (F1-score)}
	\end{minipage}
\end{figure}

\begin{figure}[ht!]
	\centering
	\begin{minipage}{0.45\textwidth}
		\centering
		\includegraphics[width=0.9\textwidth]{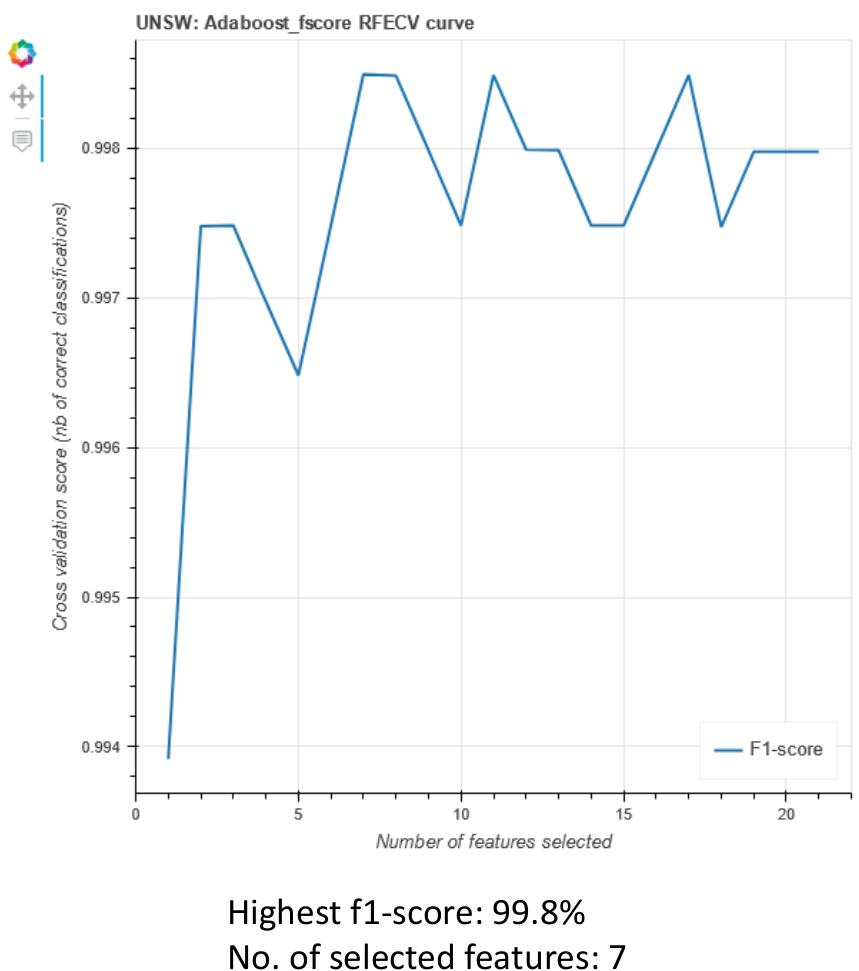} 
		\caption{RFE with Adaboost Classifier for UNSW Dataset (F1-score)}
		\label{RFE with Adaboost Classifier for UNSW Dataset (F1-score)}	
	\end{minipage}\hfill
	\begin{minipage}{0.45\textwidth}
		\centering
		\includegraphics[width=0.9\linewidth]{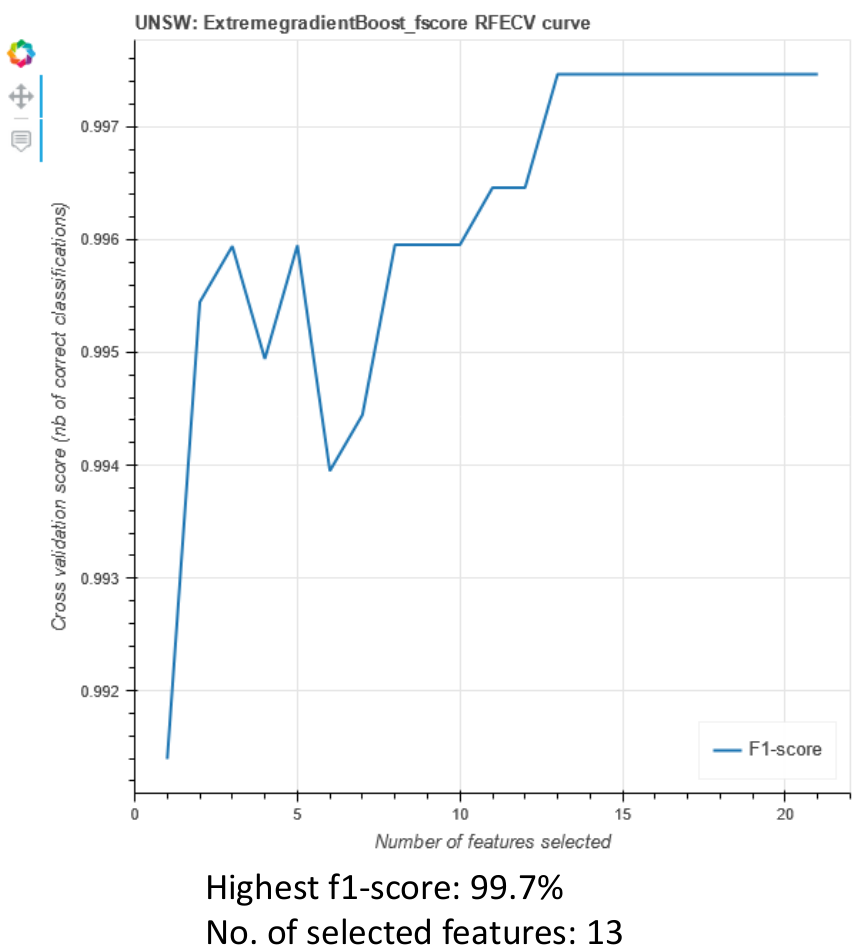}
		\caption{RFE with XGBoost Classifier for UNSW Dataset (F1-score)}
		\label{RFE with XGBoost Classifier for UNSW Dataset (F1-score)}
	\end{minipage}
\end{figure}

\begin{figure}[ht!]
	\centering
	\begin{minipage}{0.45\textwidth}
		\centering
		\includegraphics[width=0.9\textwidth]{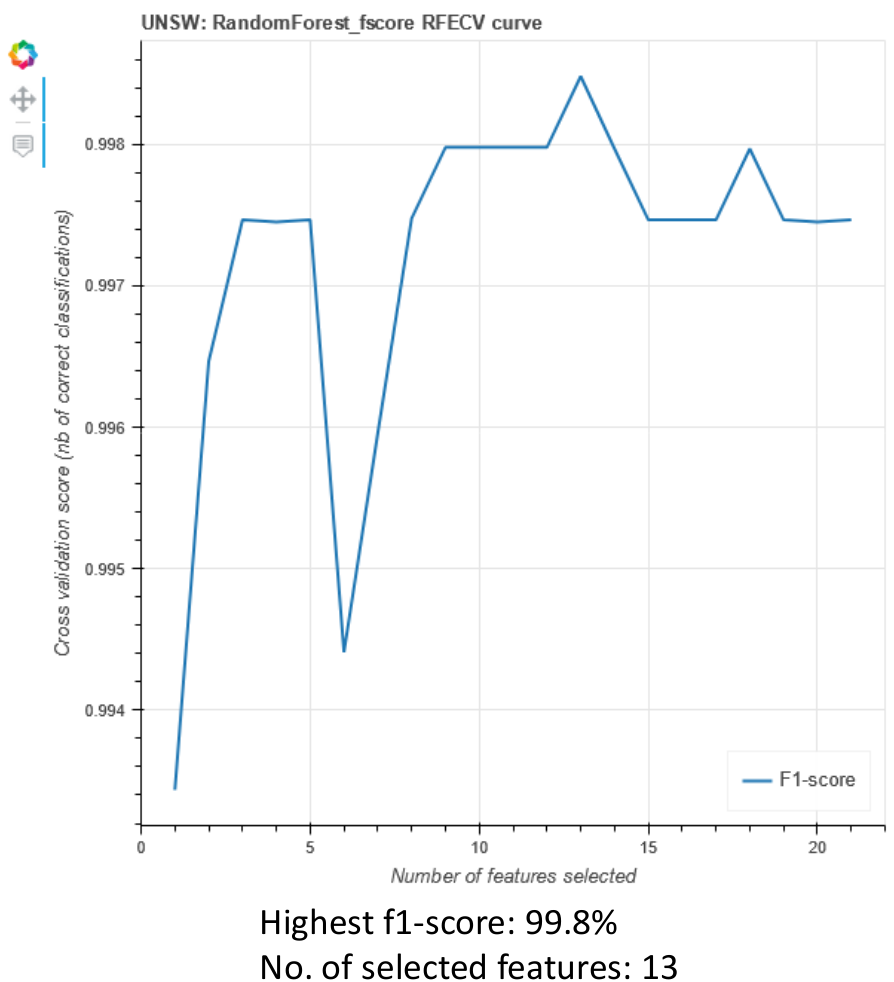} 
		\caption{RFE with Random Forest Classifier for UNSW Dataset (F1-score)}
		\label{RFE with Random Forest Classifier for UNSW Dataset (F1-score)}
	\end{minipage}\hfill
	\begin{minipage}{0.45\textwidth}
		\centering
		\includegraphics[width=0.9\textwidth]{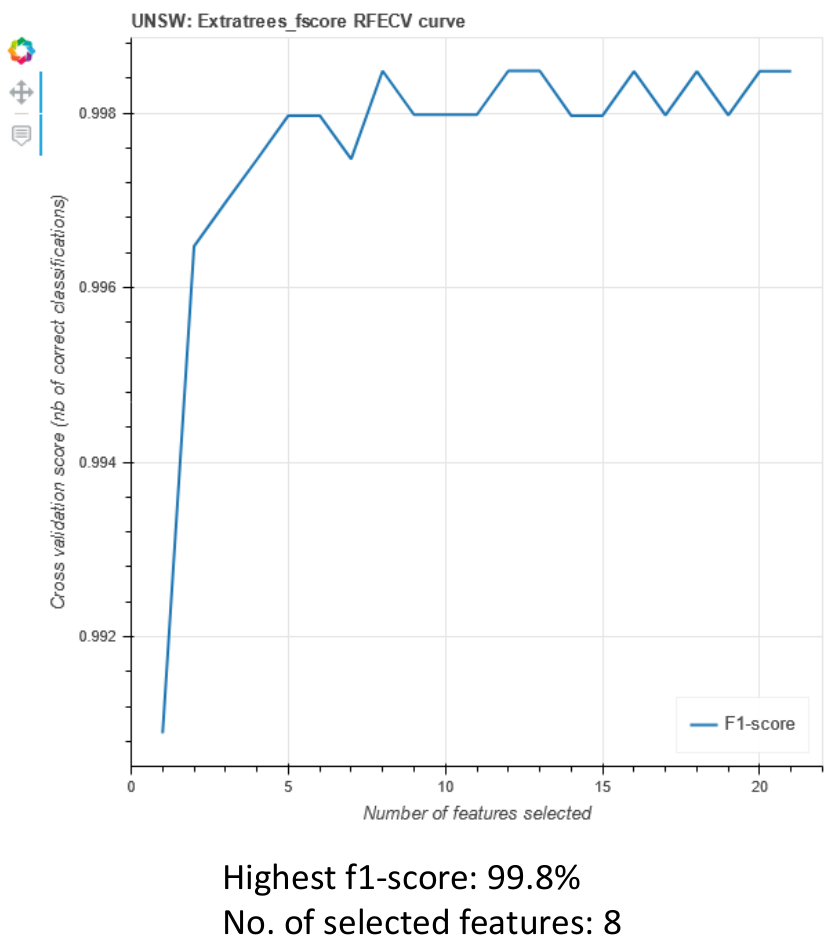} 
		\caption{RFE with Extra Trees Classifier for UNSW Dataset (F1-score)}
		\label{RFE with Extra Trees Classifier for UNSW Dataset (F1-score)}
	\end{minipage}
\end{figure}

In case of the UNSW dataset, the highest accuracy of 99.7\% is obtained by Gradient Boosting classifier with 5 features as shown in Figure \ref{RFE with Gradient Boosting Classifier for UNSW Dataset (Accuracy)}. All the other classifiers i.e AdaBoost, Extreme Gradient Boosting, Random Forest and Extra Trees show similar performance in terms of accuracy as shown in Figures \ref{RFE with Adaboost Classifier for UNSW Dataset (Accuracy)}, \ref{RFE with XGBoost Classifier for UNSW Dataset (Accuracy)}, \ref{RFE with Random Forest Classifier for UNSW Dataset (Accuracy)} and \ref{RFE with Extra Trees Classifier for UNSW Dataset (Accuracy)} respectively. However, to achieve that performance the classifiers require 7 (for Adaboost), 13 (for both Extreme Gradient Boosting and Random forest) and 8 (for Extra Trees) which is higher than the number of features required by Gradient Boosting. Hence, in this case, it is concluded that the optimal performance is given by Gradient Boosting classifier with 5 features. For the same dataset, F1-scores are illustrated in Figure \ref{RFE with Gradient Boosting Classifier for UNSW Dataset (F1-score)} for Gradient Boosting, Figure \ref{RFE with Adaboost Classifier for UNSW Dataset (F1-score)} for Adaboost, Figure \ref{RFE with XGBoost Classifier for UNSW Dataset (F1-score)} for XGBoost, Figure \ref{RFE with Random Forest Classifier for UNSW Dataset (F1-score)} for Random Forest, and Figure \ref{RFE with Extra Trees Classifier for UNSW Dataset (F1-score)} for Extra Trees classifier.

\begin{figure}[ht!]
	\centering
\begin{minipage}{0.45\textwidth}
		\centering
	\includegraphics[width=0.9\linewidth]{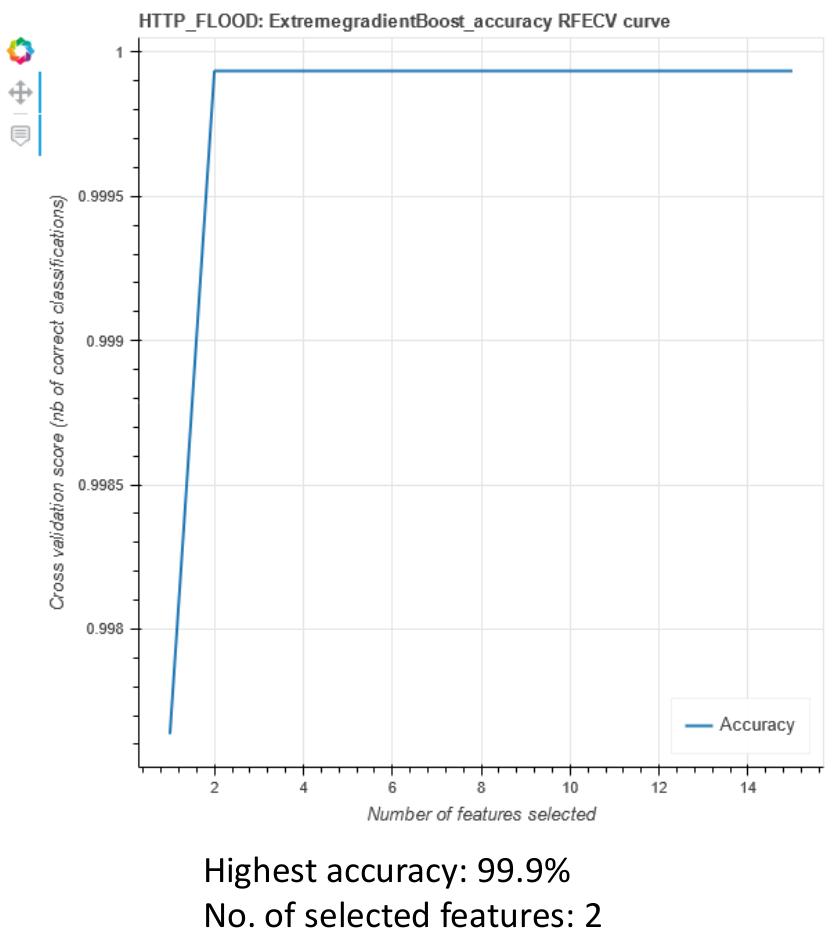}
	\caption{RFE with XGBoost Classifier for HTTP Flood Dataset (Accuracy)}
	\label{RFE with XGBoost Classifier for HTTP Flood Dataset (Accuracy)}
\end{minipage}\hfill
\begin{minipage}{0.45\textwidth}
		\centering
		\includegraphics[width=0.9\textwidth]{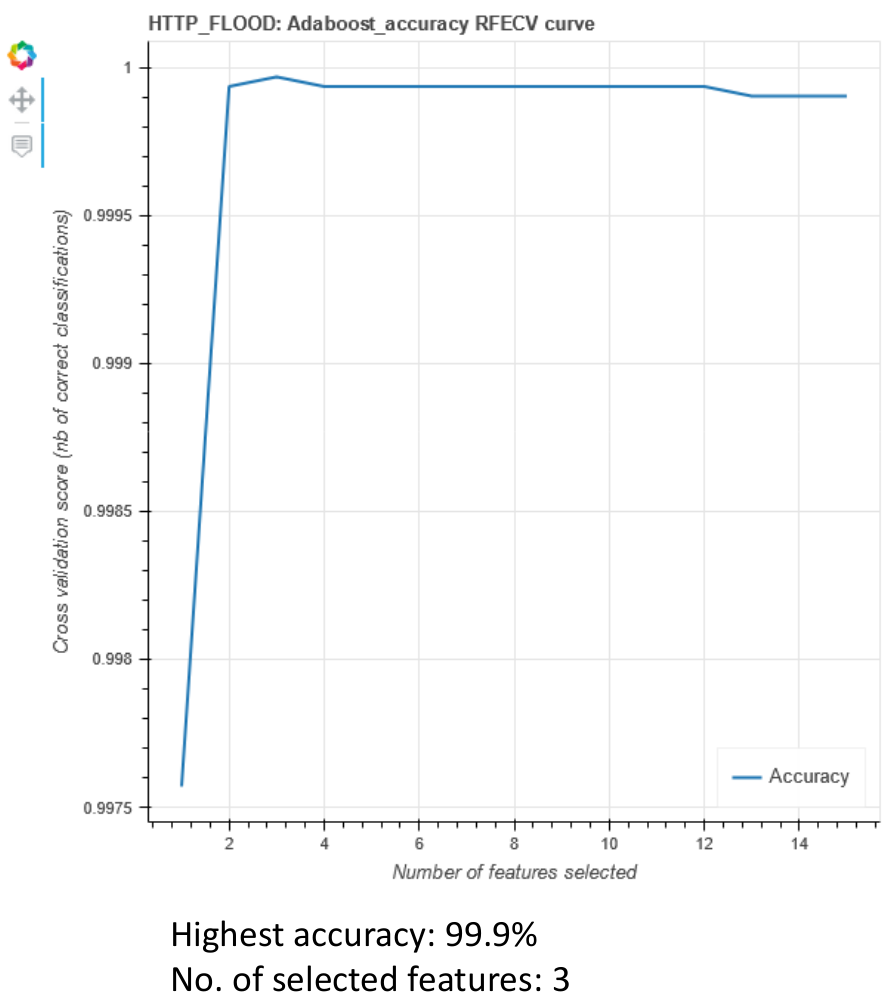} 
		\caption{RFE with Adaboost Classifier for HTTP Flood Dataset (Accuracy)}
		\label{RFE with Adaboost Classifier for HTTP Flood Dataset (Accuracy)}
	\end{minipage}
\end{figure}

\begin{figure}[ht!]
	\centering
	\begin{minipage}{0.45\textwidth}
		\centering
		\includegraphics[width=0.9\textwidth]{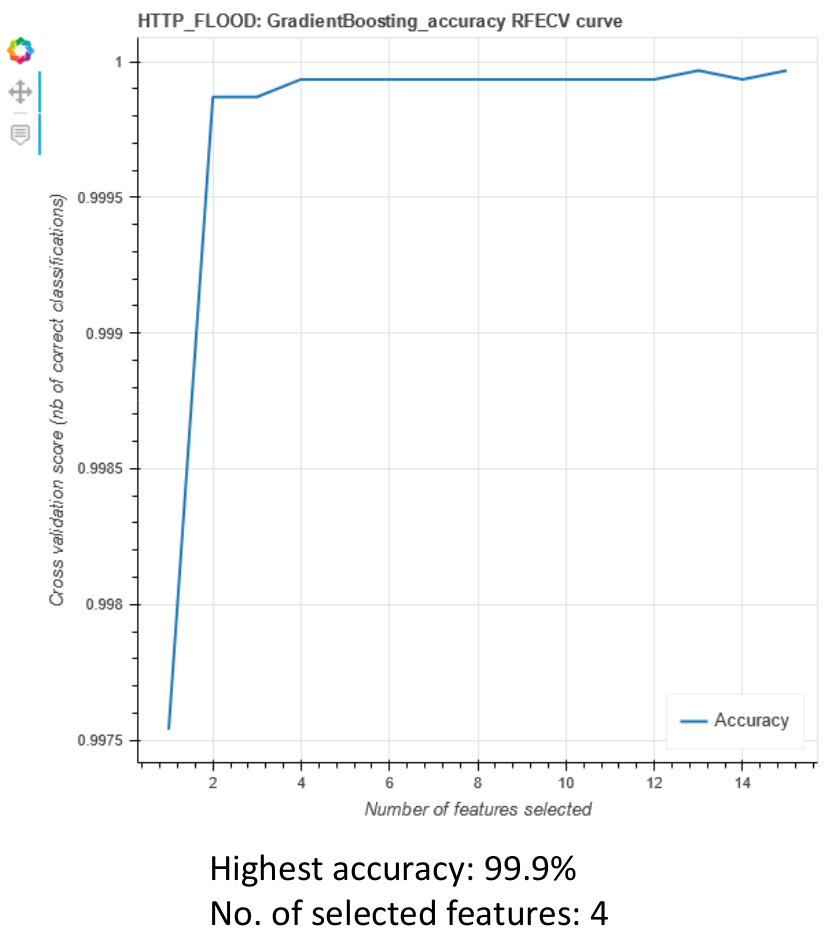} 
		\caption{RFE with Gradient Boosting Classifier for HTTP Flood Dataset (Accuracy)}
		\label{RFE with Gradient Boosting Classifier for HTTP Flood Dataset (Accuracy)}
	\end{minipage}\hfill
\begin{minipage}{0.45\textwidth}
		\centering
		\includegraphics[width=0.9\textwidth]{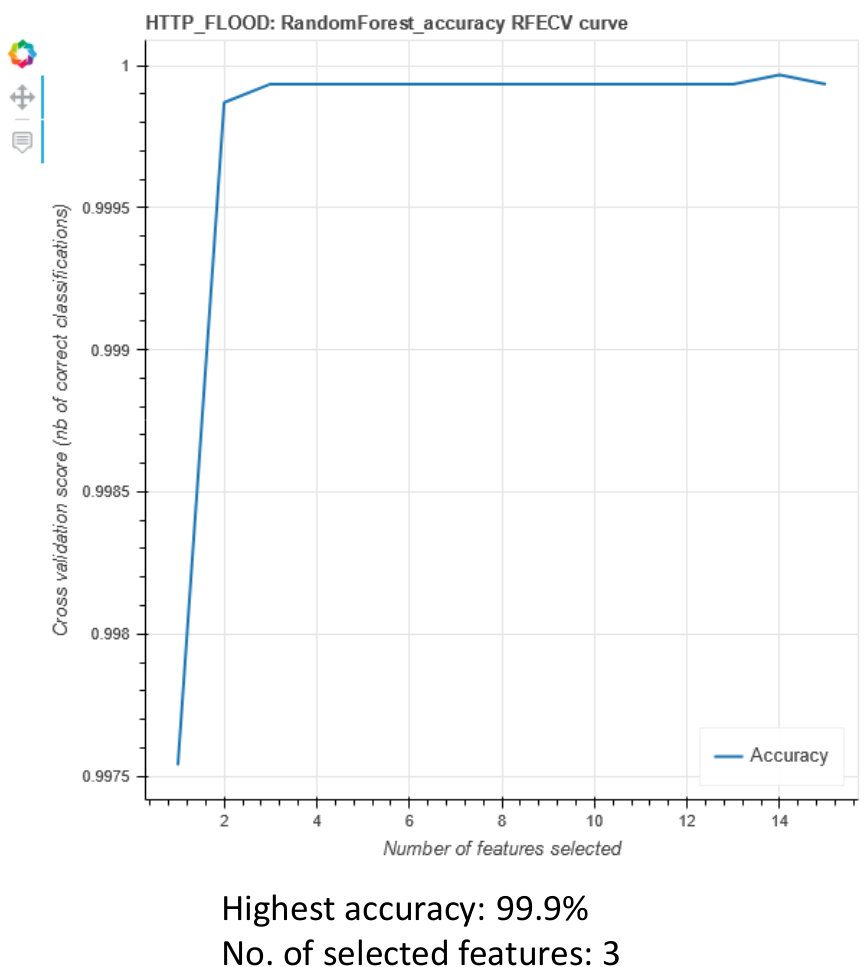} 
		\caption{RFE with Random Forest Classifier for HTTP Flood Dataset (Accuracy)}
		\label{RFE with Random Forest Classifier for HTTP Flood Dataset (Accuracy)}
	\end{minipage}
\end{figure}

\begin{figure}[ht!]
	\centering
	\begin{minipage}{0.45\textwidth}
		\centering
		\includegraphics[width=0.9\textwidth]{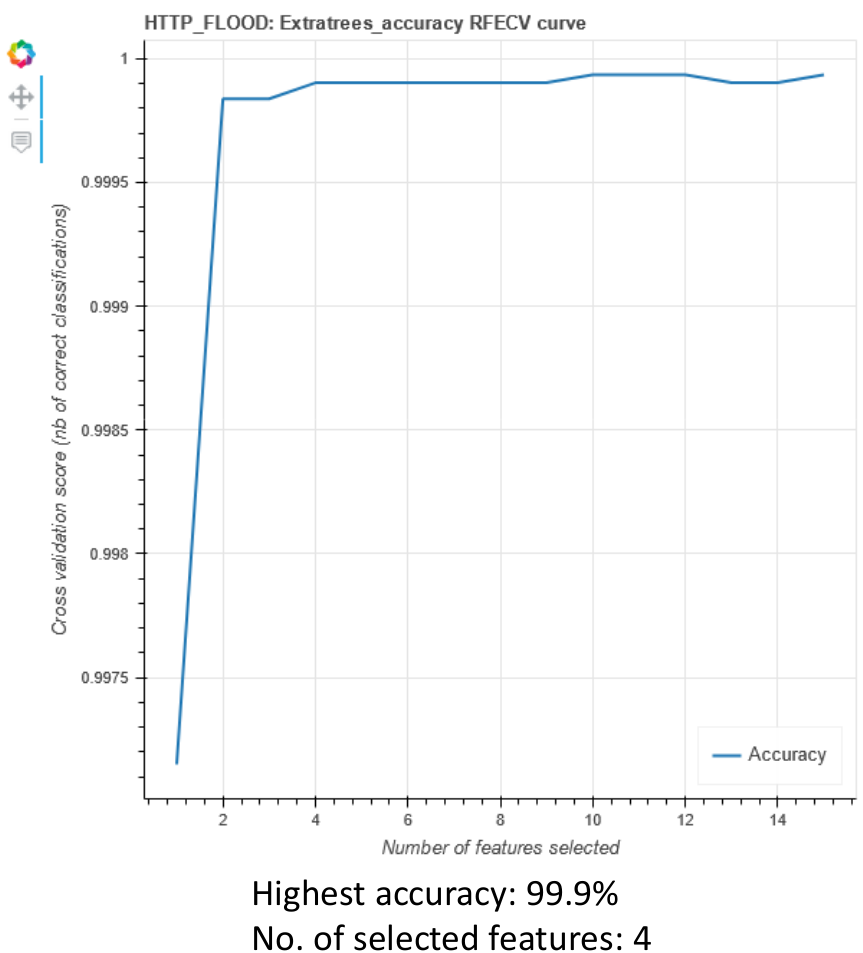} 
		\caption{RFE with Extra Trees Classifier for HTTP Flood Dataset (Accuracy)}
		\label{RFE with Extra Trees Classifier for HTTP Flood Dataset (Accuracy)}
	\end{minipage}\hfill
\begin{minipage}{0.45\textwidth}
		\centering
\includegraphics[width=0.9\textwidth]{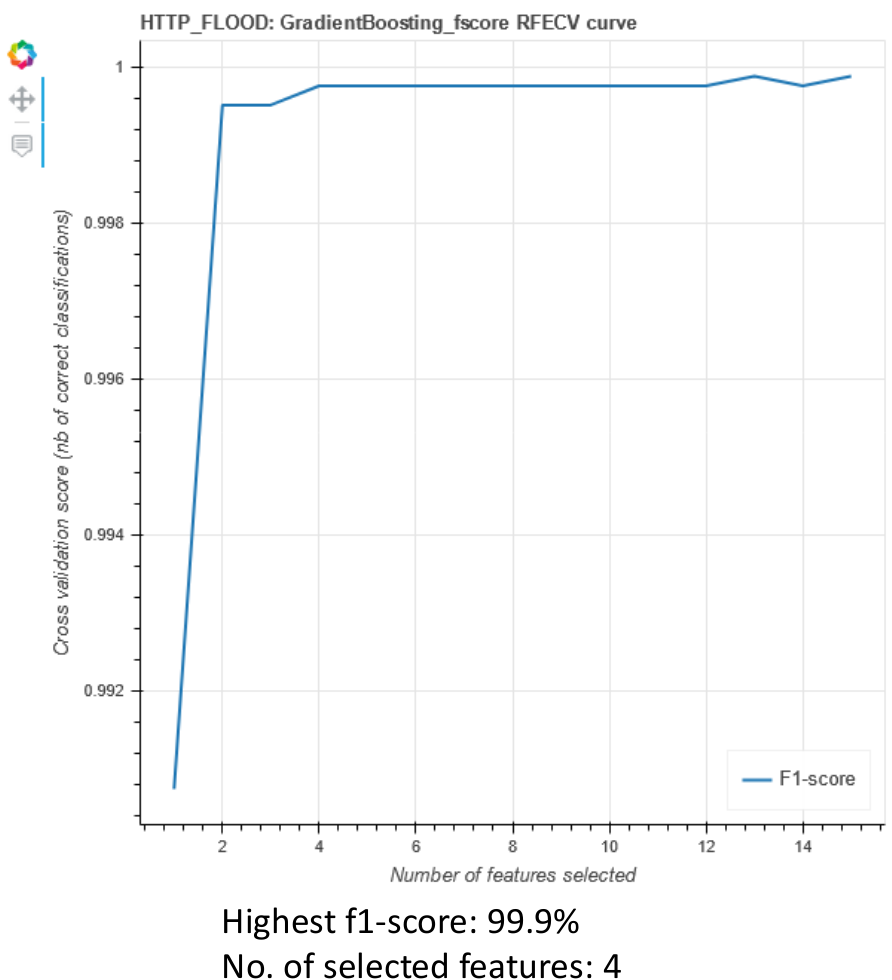} 
	\caption{RFE with Gradient Boosting Classifier for HTTP Flood Dataset (F1-score)}
	\label{RFE with Gradient Boosting Classifier for HTTP Flood Dataset (F1-score)}
\end{minipage}
\end{figure}

\begin{figure}[ht!]
	\centering
	\begin{minipage}{0.45\textwidth}
		\centering
		\includegraphics[width=0.9\textwidth]{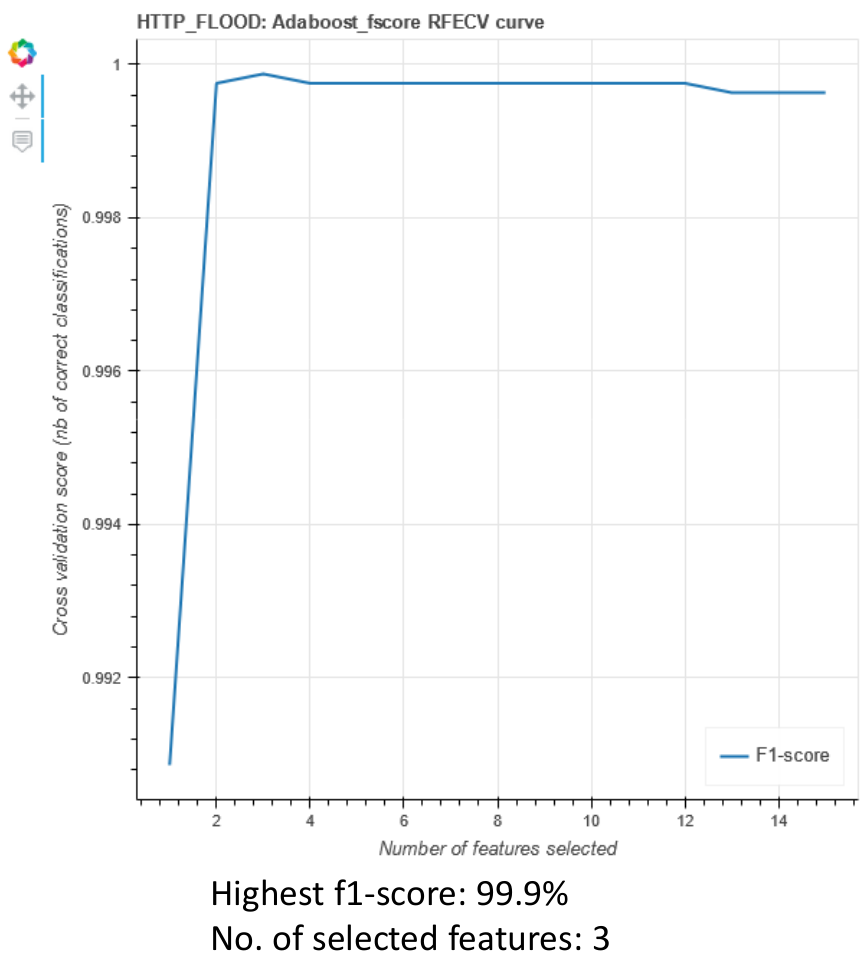} 
		\caption{RFE with Adaboost Classifier for HTTP Flood Dataset (F1-score)}
		\label{RFE with Adaboost Classifier for HTTP Flood Dataset (F1-score)}
	\end{minipage}\hfill
	\begin{minipage}{0.45\textwidth}
		\centering
		\includegraphics[width=0.9\linewidth]{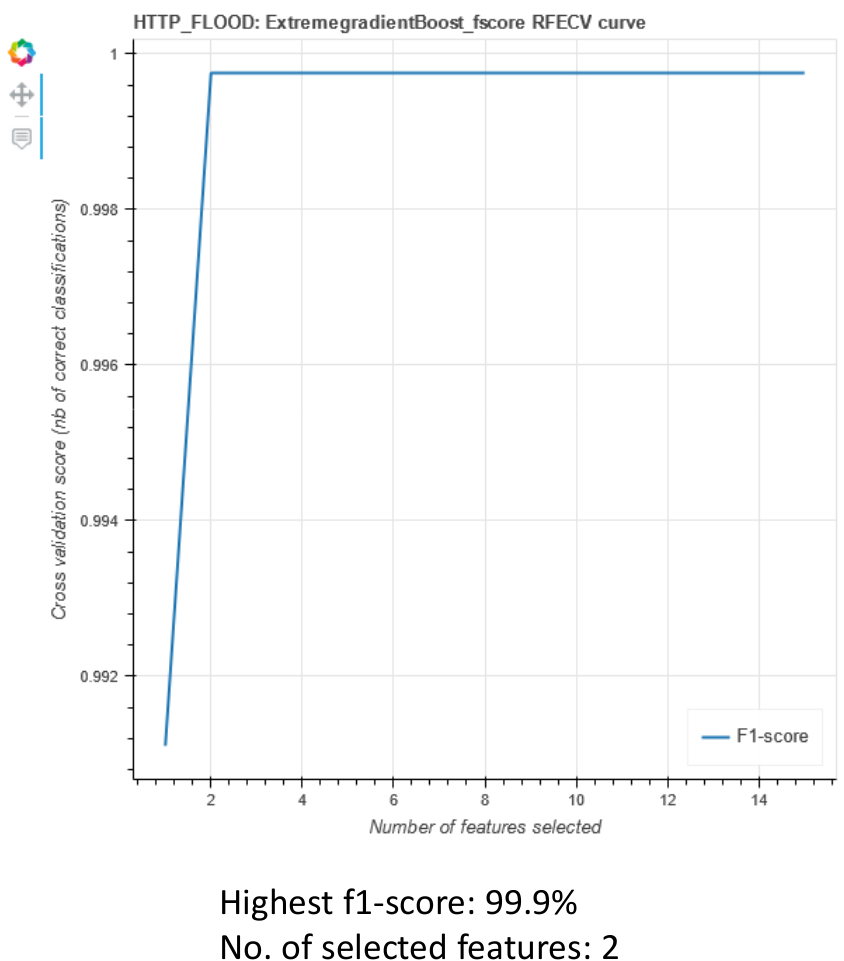}
		\caption{RFE with XGBoost Classifier for HTTP Flood Dataset (F1-score)}
		\label{RFE with XGBoost Classifier for HTTP Flood Dataset (F1-score)}
	\end{minipage}
\end{figure}

\begin{figure}[ht!]
	\centering
	\begin{minipage}{0.45\textwidth}
		\centering
		\includegraphics[width=0.9\textwidth]{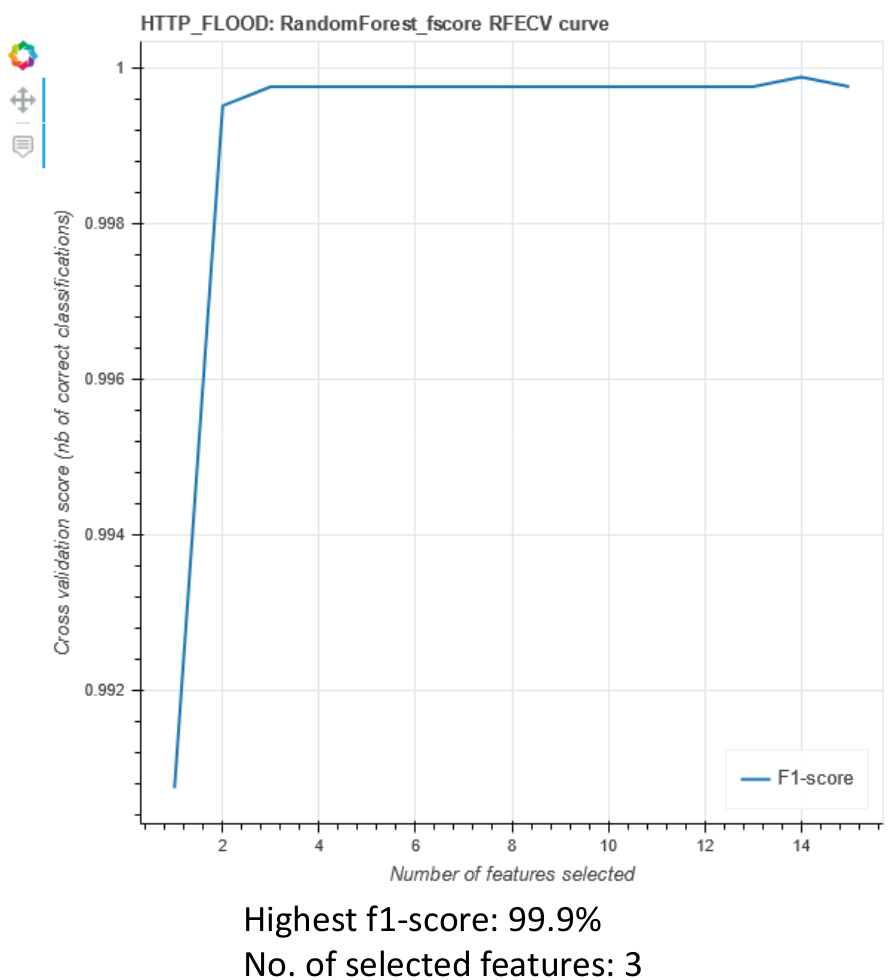} 
		\caption{RFE with Random Forest Classifier for HTTP Flood Dataset (F1-score)}
		\label{RFE with Random Forest Classifier for HTTP Flood Dataset (F1-score)}
	\end{minipage}\hfill
	\begin{minipage}{0.45\textwidth}
		\centering
		\includegraphics[width=0.9\textwidth]{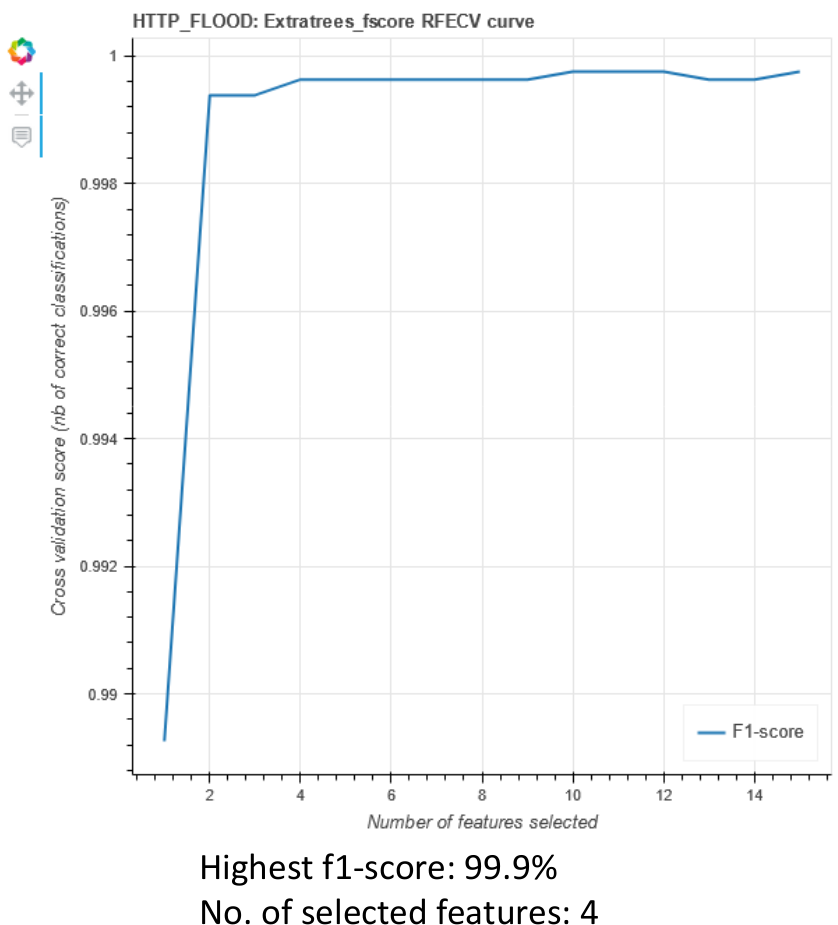} 
		\caption{RFE with Extra Trees Classifier for HTTP Flood Dataset (F1-score)}
		\label{RFE with Extra Trees Classifier for HTTP Flood Dataset (F1-score)}
	\end{minipage}
\end{figure}

On the other hand, for the HTTP Flood dataset, Extreme Gradient Boosting achieves the highest accuracy with 99.9\% with 2 features only as shown in Figure \ref{RFE with XGBoost Classifier for HTTP Flood Dataset (Accuracy)}. All other learners give similar performance with highest accuracy 99.9\% as shown in Figures \ref{RFE with Adaboost Classifier for HTTP Flood Dataset (Accuracy)} (for Adaboost), Figure \ref{RFE with Gradient Boosting Classifier for HTTP Flood Dataset (Accuracy)} (for Gradient Boosting), Figure \ref{RFE with Random Forest Classifier for HTTP Flood Dataset (Accuracy)} (for Random Forest), Figure \ref{RFE with Extra Trees Classifier for HTTP Flood Dataset (Accuracy)} (for Extra Trees). However, to achieve that performance the learners i.e. Adaboost, Gradient Boosting, Random Forest and Extra Trees require 3, 4, 3 and 4 features respectively which is higher than the number of features required by Extreme Gradient Boosting classifier. Hence, in this case, it is concluded that the optimal performance is given by Extreme Gradient Boosting classifier with 2 features. For the same dataset, F1-scores are illustrated in Figure \ref{RFE with Gradient Boosting Classifier for HTTP Flood Dataset (F1-score)} for Gradient Boosting, Figure \ref{RFE with Adaboost Classifier for HTTP Flood Dataset (F1-score)} for Adaboost, Figure \ref{RFE with XGBoost Classifier for HTTP Flood Dataset (F1-score)} for XGBoost, Figure \ref{RFE with Random Forest Classifier for HTTP Flood Dataset (F1-score)} for Random Forest, and Figure \ref{RFE with Extra Trees Classifier for HTTP Flood Dataset (F1-score)} for Extra Trees classifier. 


Table \ref{Top 10 Ranked Features of the HTTP Flooding Datasets} shows the top 10 ranked features for each of the HTTP Flooding dataset considered as given by the proposed incremental feature selection method. The columns \textit{feature name} and \textit{index number} gives the name of the feature and the index number in the dataset. 

\begin{table}[ht!]
	\centering
	\caption{Top 10 Ranked Features of the HTTP Flooding Datasets}
	\label{Top 10 Ranked Features of the HTTP Flooding Datasets}
	\resizebox{\textwidth}{!}{%
		\begin{tabular}{@{}lllllll@{}}
			\toprule

			& \multicolumn{2}{c}{UNSW} & \multicolumn{2}{c}{CICIDS} & \multicolumn{2}{c}{HTTP-FLOOD} \\ \midrule
			& Feature Name & \begin{tabular}[c]{@{}l@{}}Index \\ Number\end{tabular} & Feature Name & \begin{tabular}[c]{@{}l@{}}Index \\ Number\end{tabular} & Feature Name & \begin{tabular}[c]{@{}l@{}}Index \\ Number\end{tabular} \\ \midrule
			1 & dport & 3 & Flow Bytes/s & 14 & SEQ\_NUMBER & 8 \\
			2 & daddr & 2 & Bwd Packets/s & 37 & NODE\_NAME\_FROM & 11 \\
			3 & drate & 15 & Init\_Win\_bytes\_backward & 67 & FIRST\_PKT\_SENT & 24 \\
			4 & AR\_P\_Proto\_P\_SrcIP & 22 & Destination Port & 0 & BYTE\_RATE & 18 \\
			5 & N\_IN\_Conn\_P\_SrcIP & 25 & Init\_Win\_bytes\_forward & 66 & LAST\_PKT\_RESEVED & 25 \\
			6 & srate & 14 & Packet Length Mean & 40 & PKT\_DELAY & 21 \\
			7 & dur & 8 & Average Packet Size & 52 & FID & 7 \\
			8 & TnP\_Per\_Dport & 21 & Max Packet Length & 39 & NODE\_NAME\_TO & 12 \\
			9 & bytes & 5 & Avg Bwd Segment Size & 54 & NUMBER\_OF\_BYTE & 10 \\
			10 & TnBPSrcIP & 16 & Bwd Packet Length Mean & 12 & UTILIZATION & 20 \\ \bottomrule
		\end{tabular}%
	}
\end{table}

\subsection{Comparison with other Feature Selection Methods}
The proposed incremental feature selection method which is designed to detect HTTP Flood attacks in the application layer is compared against seven state-of-the-art feature selection methods such as MIFS \cite{battiti1994using}, CMIM \cite{fleuret2004fast}, and mRMR \cite{peng2005feature}, DISR \cite{brown2012conditional}, JMI, Gini index and ANOVA feature selection. Figure \ref{F1-score Comparison for CICIDS Dataset}, \ref{F1-score Comparison for HTTP-Flood Dataset} and \ref{F1-score Comparison for UNSW Dataset} illustrates the comparative analysis of INFS-MICC against its counter parts. For comparative analysis, F1-score is used as an evaluation metric as it is a better measure than accuracy in case of unbalanced datasets. For the CICIDS dataset, since the proposed method obtained highest accuracy with 3 features, the F1-score comparison is also done considering 3 features only for each feature selection method. Similarly, for UNSW and HTTP-Flood dataset highest accuracies are obtained with 5 and 2 features respectively. Hence, comparison for these two datasets are done considering the said number of features for each feature selection method. From the comparative analysis, it is seen that the proposed method gives on par or better performance compared to the other feature selection methods.

\section{Conclusion}
INFS-MICC is an incremental feature selection method, proposed to detect HTTP-Flooding attacks. The proposed method aids to identify a final ranked list of feature subset which consists of highly relevant and irredundant features. For computing the relevance, feature-class mutual information is considered and for the irredundancy among features, the feature-feature correlation is computed. The main highlight of our method is that it is incremental in nature, as it can handle added-in data which avoids re-computation of the whole dataset. The proposed method is evaluated with three HTTP-Flooding datasets and five ensemble predictors using two evaluation metrics namely Accuracy and F1-score. For two out of the three datasets Extreme Gradient Boosting gives optimal performance whereas for one of the dataset Gradient Boosting classifier performs the best.

\bibliographystyle{unsrt}  
\bibliography{samplepaper}

\end{document}